\documentclass[fp,twocolumn]{jpsj3}
\usepackage{txfonts}
\usepackage{bm}
\usepackage{color}
\usepackage{graphicx}
\usepackage{amsmath}
\usepackage{amsmath,amssymb}
\usepackage{mathrsfs}
\def\agt{>\kern -9truept\lower 4.0truept\hbox{$\displaystyle\sim$}}
\def\alt{<\kern -9truept\lower 4.0truept\hbox{$\displaystyle\sim$}}

\title{Raman Characterization of Two-Dimensional Quasiperiodic Antiferromagnets
       on Various Lattices: Spin-Orbit Mechanism}

\author{Takashi Inoue and Shoji Yamamoto\thanks{yamamoto@phys.sci.hokudai.ac.jp}}

\inst{Department of Physics, Hokkaido University, Sapporo 060-0810, Japan} 

\abst{
We study first-order (single-magnon) inelastic light scatterings in spin-$\frac{1}{2}$
two-dimensional quasiperiodic antiferromagnets in comparison with those emergent on
periodic lattices.
Unlike second-order (two-magnon) Raman scatterings based on an exchange interaction between
neighboring spins, the present observations involve an indirect electric-dipole coupling
which proceeds through a spin-orbit interaction.
We discuss the nearest-neighbor antiferromagnetic \textit{XXZ} Hamiltonian on various
quasiperiodic and periodic bipartite lattices.
The first-order spectra, of our present interest, consist only of rotation-invariant and
mirror-symmetric magnons, while the second-order ones cannot select any particular magnon.
With the exchange anisotropy moving away from the Ising limit toward the Heisenberg isotropic
point, every initial delta-function peak bifurcates or divides into more in each individual manner
on quasiperiodic lattices,
while it remains singly peaked all the way on periodic lattices.
Such splittings depend on how many types of isocoordinated sites for each coordination number
and relative positions between those of the same type.
A perpendicular-space representation of the first-order Raman spectrum serves as a fingerprint
of each quasiperiodic tiling.
}

\begin{document}
\maketitle

\section{Introduction}
   Quasicrystals are exotic materials that lack translational symmetry but exhibit
long-range order \cite{L2477,L596}.
Since their first discovery in aluminum-manganese alloys \cite{S1951}, a wide variety of
quasicrystalline materials have been synthesized
\cite{TL1505,TL1587,T537,T58,D1013,K154,T19938,Y247}.
While early explorations were limited to synthetic intermetallic compounds \cite{T5352},
recent studies have reported  quasicrystals in diverse contexts,
including natural minerals such as
icosahedral $\mathrm{Al_{63}Cu_{24}Fe_{13}}$ \cite{B1306,B928} and decagonal
$\mathrm{Al_{71}Ni_{24}Fe_{5}}$ \cite{B9111} quasicrystals
and soft-matter systems such as supramolecular dendritic liquid \cite{D1197,Z157} and
polymeric quasicrystals \cite{H195502}.
It deserves special mention that a quasiperiodic structure can be fabricated on a
$30^{\circ}$-twisted bilayer graphene as well \cite{A782,M165430,T4583,Y6928,P3313}.

   Recent progress in optical-lattice techniques has enabled the creation of artificial
quasiperiodic potentials \cite{G3363,V110404,S200604}.
Two-dimensional quasiperiodic optical potentials with fivefold \cite{SP053607} and
eightfold \cite{J66003,J149} rotational symmetries have been theoretically designed
in terms of standing-wave lasers.
They have also been experimentally demonstrated through observation of
the matter-wave diffraction of trapped atoms \cite{SP053607,V110404},
the superfluid-insulator transition of Bose-Einstein condensates \cite{S200604}, and
the Bose glass \cite{Y338}.
Ultracold atoms in optical lattices are employed as a simulator of condensed matter
physics, thanks to the flexibility of lattice geometries and the tunability of system
parameters such as hopping amplitudes and interatomic interactions \cite{B885}.
These developments have made it feasible to emulate diverse theoretical models
in quasiperiodic systems, including tight-binding models
\cite{K2740,A1621,M064213,O014204,A064207}, fermionic \cite{K214402,K115125,K360} and
bosonic \cite{G224201,H114005,J053609} Hubbard models, and Heisenberg models
\cite{W177205,W104427,J212407,S104427}.
By assigning two internal states of trapped atoms to effective spins and controlling
virtual spin-dependent tunneling energies, one can engineer
effective spin models with tunable exchange anisotropy in the strongly
interacting Mott-insulating regime \cite{D090402}.

   Magnetic Raman scattering \cite{M490}, an inelastic two-photon process mediated by
spin-dependent electric polarizability, provides a powerful probe of magnetic excitations.
The microscopic mechanisms of magnetic Raman scattering in antiferromagnets include
(i) processes based on the exchange mechanism \cite{F514,S1068,S365}, involving virtual
electron hopping induced by light, and (ii) processes based on the spin-orbit (SO) mechanism
\cite{E189,F514,K2993}, involving orbital excitations induced by electric-dipole interaction.
   The exchange-mechanism Raman scattering is mediated by the Brillouin-zone-edge
second-order (two-magnon) excitations \cite{F514,K4418}, enabling accurate determination
of exchange constants \cite{BR11930}.
The second-order spectra have been investigated in two-dimensional
quasiperiodic Heisenberg antiferromagnets \cite{I053701,I2000118}.
At the lowest order of perturbation, pair-exchange excitations give rise exclusively to
the $\mathrm{E}_{2}$ symmetry, which is the one and only Raman-active mode, thereby
eliminating any dependence on linear incident and scattered polarizations. 
At higher perturbative orders, dynamic ring-exchange and chiral spin fluctuations activate
additional symmetry species $\mathrm{A}_{1}$ and $\mathrm{A}_{2}$, whose polarization
dependencies have been clarified through group-theoretical analysis.
In contrast, the SO-mechanism Raman scattering is mediated by Brillouin-zone-center
first-order (single-magnon) excitations \cite{F514,L6009,C641}, allowing precise probing of
anisotropy gaps \cite{G027001,SN140401R,B024416}.
The SO and exchange Raman responses can be distinguished by their contrasting
polarization dependence, with the former depending on the direction of the spin easy axis
and the latter not \cite{B024416}.

   Ultracold fermionic atoms in quasiperiodic optical lattices may possibly visualize
SO-mechanism Raman scattering of our interest.
Standing-wave lasers \cite{SP053607,J66003,J149} can be used to realize and tune artificial
spin-orbit couplings in both fermionic and bosonic systems \cite{L83, D1523, C095302}.
Antiferromagnetic correlations between ultracold atoms have indeed been detected
via off-resonant light scattering \cite{A031036} and spin-sensitive Bragg
scattering \cite{H211} and can be formulated in terms of Raman spectroscopy as well
\cite{B063618}.
Thus motivated, we study the SO-mechanism first-order Raman spectra for
the spin-$\frac{1}{2}$ antiferromagnetic \textit{XXZ} model on
various two-dimensional quasiperiodic bipartite lattices in comparison with periodic ones.

\section{Two-dimensional Quasiperiodic Lattice}
   We investigate three two-dimensional quasiperiodic lattices: the Penrose, Ammann-Beenker,
 and Socolar lattices.
Each of these lattices is bipartite, as it is composed of even-edged prototiles.
No sublattice imbalance arises in the thermodynamic limit \cite{K214402,K115125,K360}.
For all three lattices, the spatial dimension $d$ is $2$, and the index dimension
(or rank) $D$ is $4$.
In our calculations, we employ finite open clusters with $\mathbf{C}_{5\mathrm{v}}$,
$\mathbf{C}_{8\mathrm{v}}$, and $\mathbf{C}_{3\mathrm{v}}$ point symmetries for the
Penrose, Ammann-Beenker, and Socolar lattices, respectively.

\begin{figure}[tb]
\centering
\includegraphics[width=\linewidth]{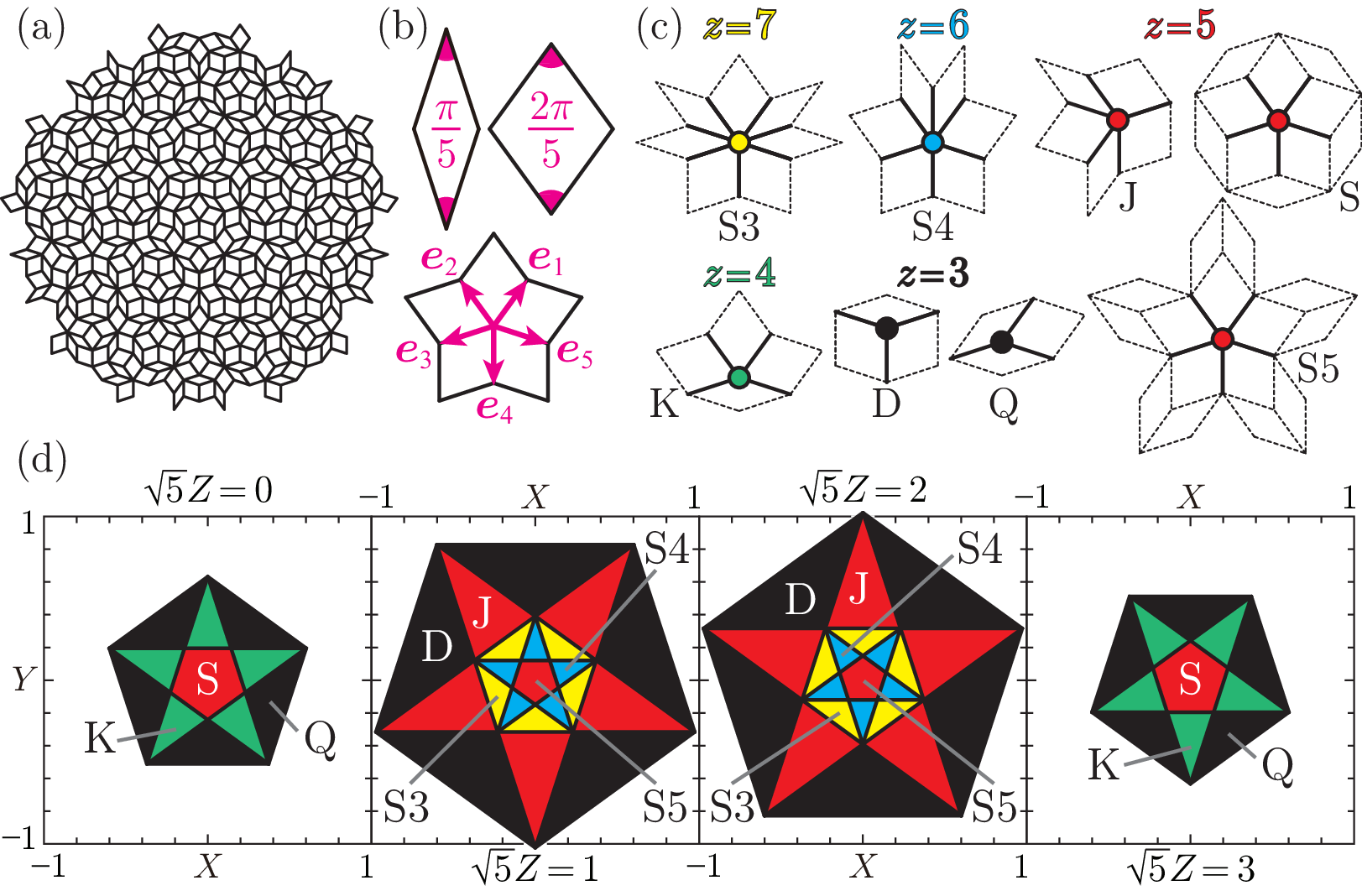}
\caption{%
         (a) $\mathbf{C}_{5\mathrm{v}}$-symmetric finite Penrose lattice with $L=526$ sites.
         (b) The Penrose lattice is generated from two rhombic prototiles with acute angles
             of $\pi/5$ and $2\pi/5$.
             The canonical basis vectors of a five-dimensional hypercubic lattice are
             projected onto five vectors, denoted $\bm{e}_{1}$, $\bm{e}_{2}$, $\bm{e}_{3}$,
             $\bm{e}_{4}$, and $\bm{e}_{5}$.
             Any four of these vectors can be chosen as the primitive translation vectors
             for the two-dimensional Penrose lattice, with the constraint
             $\sum_{n=1}^{5}\bm{e}_{n}=\bm{0}$.
         (c) Eight types of vertices in the Penrose lattice with coordination numbers
             $z=3$ to $7$. 
         (d) The projection windows of the Penrose lattice consist of a three-dimensional
             stack of four pentagons located at $Z=0$, $1/\sqrt{5}$, $2/\sqrt{5}$,
             and $3/\sqrt{5}$.
             The colors of the domains correspond to the vertices shown in (c).
         }
\label{F:PenroseLattice}
\end{figure}

\subsection{Penrose lattice}
   A two-dimensional Penrose lattice \cite{P266} [Fig.~\ref{F:PenroseLattice}(a)] is
constructed from two rhombic prototiles with acute angles of $\frac{\pi}{5}$ and 
$\frac{2\pi}{5}$ [Fig.~\ref{F:PenroseLattice}(b)].
The Penrose lattice is obtained as a projection of a five-dimensional hypercubic lattice
onto a two-dimensional real space,
\begin{align}
   &
   \left[
   \begin{array}{c}
     x
    \\
     y
    \\
     X
    \\
     Y
    \\
     Z
   \end{array}
   \right]
  =
   \sqrt{\frac{2}{5}}
   \left[
   \begin{array}{ccccc}
     c_{0} & c_{1} & c_{2} & c_{3} & c_{4}
    \\
     s_{0} & s_{1} & s_{2} & s_{3} & s_{4}
    \\
     c_{0} & c_{2} & c_{4} & c_{6} & c_{8}
    \\
     s_{0} & s_{2} & s_{4} & s_{6} & s_{8}
    \\
     \frac{1}{\sqrt{2}} & \frac{1}{\sqrt{2}} & \frac{1}{\sqrt{2}} & \frac{1}{\sqrt{2}} & \frac{1}{\sqrt{2}}
   \end{array}
   \right]
   \left[
   \begin{array}{c}
     m_{1}
    \\
     m_{2}
    \\
     m_{3}
    \\
     m_{4}
    \\
     m_{5}
   \end{array}
   \right];
   \allowdisplaybreaks
   \nonumber \\
   &
   c_{n}=\cos\left(\frac{2\pi}{5}n+\frac{\pi}{10}\right),
  \ 
   s_{n}=\sin\left(\frac{2\pi}{5}n+\frac{\pi}{10}\right),
  \ 
   m_{n}\in\mathbb{Z},
\label{E:projectionPenrose}
\end{align}
where $(x,y)$ and $(X,Y,Z)$ are the coordinates in the two-dimensional real and
three-dimensional perpendicular spaces, respectively.
The prefactor $\sqrt{2/5}$ serves as the lattice constant.
A one-to-one correspondence exists between the positions in the real space and those
in the perpendicular space.
Every vertex of the Penrose lattice is represented by an integer linear combination of
four independent primitive lattice vectors [Fig.~\ref{F:PenroseLattice}(b)].
The infinite Penrose lattice consists of eight types of local vertices, whose coordination
numbers $z$ range from 3 to 7 [Fig.~\ref{F:PenroseLattice}(c)] \cite{dB39,dB53}.
In the three-dimensional perpendicular space of the Penrose lattice, the projection window
consists of a stack of four pentagons located at $Z=0$, $1/\sqrt{5}$, $2/\sqrt{5}$, and
$3/\sqrt{5}$.
The vertex sites contained in $Z=0$ and $Z=2/\sqrt{5}$ layers form the even sublattice,
while those in $Z=1/\sqrt{5}$ and $Z=3/\sqrt{5}$ layers form the odd sublattice.
The pentagonal projection windows are partitioned into compact regions, each corresponding
to one of the eight vertex types [Fig.~\ref{F:PenroseLattice}(d)].
Their areas and structures can be categorized into two classes, symmetric with respect to
the even and odd sublattices.
It suffices to consider only the set associated with a single sublattice in the
perpendicular-space analysis.

   Two tilings are said to be \textit{locally isomorphic} if every finite patch contained in
either tiling can be found in the other as well \cite{L596,S617}.
The set of all tilings locally isomorphic to a given tiling is the local isomorphism (LI)
class of the tiling.
In this sense, the infinite Penrose lattice has decagonal symmetry \cite{S10519},
since it belongs to the same LI class before and after symmetry operations.

\begin{figure}[tb]
\centering
\includegraphics[width=0.85\linewidth]{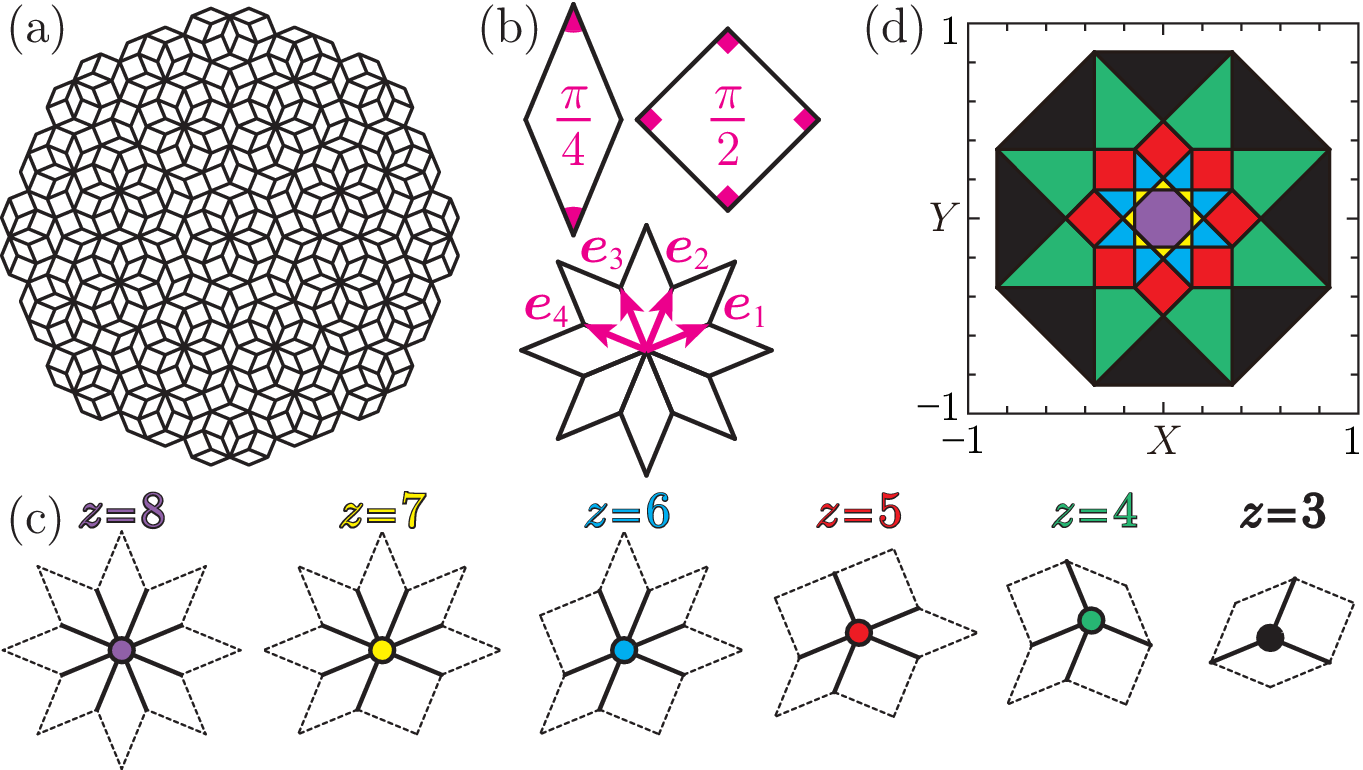}
\caption{%
         (a) $\mathbf{C}_{8\mathrm{v}}$-symmetric finite Ammann-Beenker lattice with
             $L=481$ sites.
         (b) The Ammann-Beenker lattice is generated from a rhombus with an acute angle
             of $\pi/4$ and a square.
             The canonical basis vectors of a four-dimensional hypercubic lattice are
             projected onto four vectors, denoted $\bm{e}_{1}$, $\bm{e}_{2}$, $\bm{e}_{3}$,
             and $\bm{e}_{4}$, which serve as the primitive translation vectors for the
             two-dimensional Ammann-Beenker lattice.
         (c) Six types of vertices in the Ammann-Beenker lattice with coordination numbers
             $z=3$ to $8$. 
         (d) The projection window of the Ammann-Beenker lattice is a single octagon.
             The colors of the domains correspond to the vertices shown in (c).
         }
\label{F:ABLattice}
\end{figure}

\subsection{Ammann-Beenker lattice}
   A two-dimensional Ammann-Beenker lattice \cite{S10519,B8091} [Fig.~\ref{F:ABLattice}(a)]
is constructed from a rhombus with an acute angle of $\frac{\pi}{4}$ and a square
[Fig.~\ref{F:ABLattice}(b)].
The Ammann-Beenker lattice is obtained as a projection of a four-dimensional hypercubic
lattice onto a two-dimensional real space,
\begin{align}
   &
   \left[
   \begin{array}{c}
     x
    \\
     y
    \\
     X
    \\
     Y
   \end{array}
   \right]
  =
   \frac{1}{\sqrt{2}}
   \left[
   \begin{array}{cccc}
     c_{0} & c_{1} & c_{2} & c_{3}
    \\
     s_{0} & s_{1} & s_{2} & s_{3}
    \\
     c_{0} & c_{3} & c_{6} & c_{9}
    \\
     s_{0} & s_{3} & s_{6} & s_{9}
   \end{array}
   \right]
   \left[
   \begin{array}{c}
     m_{1}
    \\
     m_{2}
    \\
     m_{3}
    \\
     m_{4}
   \end{array}
   \right];
   \allowdisplaybreaks
   \nonumber \\
   &
   c_{n}=\cos\left(\frac{\pi}{4}n\right),
  \ 
   s_{n}=\sin\left(\frac{\pi}{4}n\right),
  \ 
   m_{n}\in\mathbb{Z},
\label{E:projectionAB}
\end{align}
where $(x,y)$ and $(X,Y)$ are the coordinates in the two-dimensional real and perpendicular
spaces.
The prefactor $1/\sqrt{2}$ serves as the lattice constant.
Every vertex of the Ammann-Beenker lattice is represented by an integer linear combination
of four independent primitive lattice vectors [Fig.~\ref{F:ABLattice}(b)].
The infinite Ammann-Beenker lattice consists of six types of local vertices with $z$
ranging from 3 to 8 [Fig.~\ref{F:ABLattice}(c)].
The projection window in the two-dimensional perpendicular space of the Ammann-Beenker
lattice is a regular octagon, where, unlike the Penrose lattice, the two sublattices occupy
the same window.
The octagonal projection window is partitioned into compact regions, each corresponding to
one of the six vertex types [Fig.~\ref{F:ABLattice}(d)].

\begin{figure}[tb]
\centering
\includegraphics[width=\linewidth]{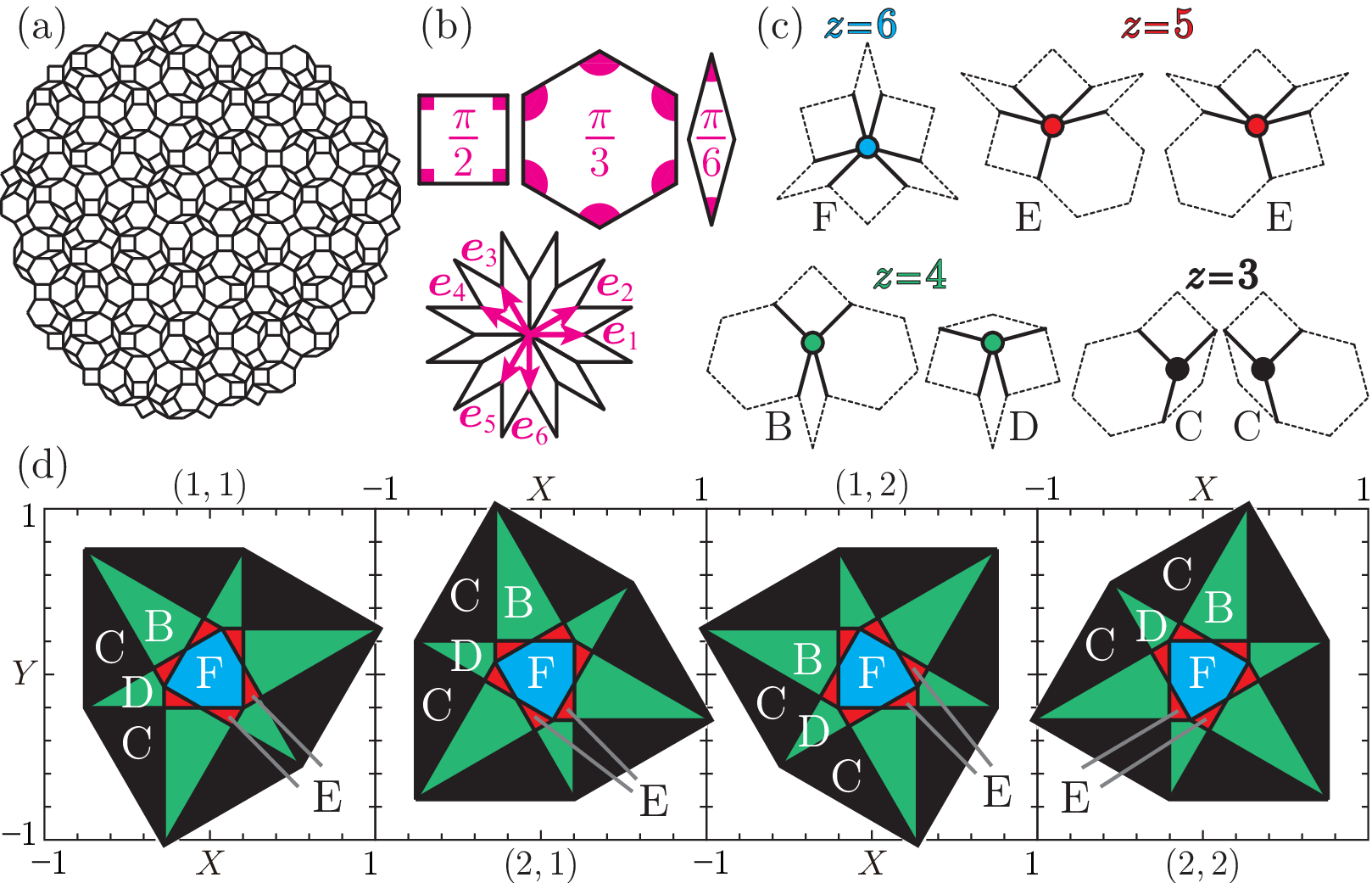}
\caption{%
         (a) $\mathbf{C}_{3\mathrm{v}}$-symmetric finite Socolar lattice with $L=784$ sites.
         (b) The Socolar lattice is generated from a rhombus with an acute angle of $\pi/6$,
             a regular hexagon, and a square.
             The canonical basis vectors of a six-dimensional hypercubic lattice are
             projected onto six vectors, denoted $\bm{e}_{1}$, $\bm{e}_{2}$, $\bm{e}_{3}$,
             $\bm{e}_{4}$, $\bm{e}_{5}$, and $\bm{e}_{6}$.
             Two vectors chosen from $\{\bm{e}_{1}, \bm{e}_{3}, \bm{e}_{5}\}$ and two from
             $\{\bm{e}_{2}, \bm{e}_{4}, \bm{e}_{6}\}$ form the primitive translation vectors
             of the two-dimensional Socolar lattice, with the constraints
             $\bm{e}_{1}+\bm{e}_{3}+\bm{e}_{5}=\bm{0}$ and
             $\bm{e}_{2}+\bm{e}_{4}+\bm{e}_{6}=\bm{0}$.
         (c) Five types of vertices in the Socolar lattice with coordination numbers
             $z=3$ to $6$. 
         (d) The projection window of the Socolar lattice consists of four hexagons
             specified by $(\sqrt{3}i,\sqrt{3}j)=(1,1)$, $(1,2)$, $(2,1)$, and $(2,2)$.
             The colors of the domains correspond to the vertices shown in (c).
         }
\label{F:SocolarLattice}
\end{figure}

\subsection{Socolar lattice}
   A two-dimensional Socolar lattice \cite{S10519} [Fig.~\ref{F:SocolarLattice}(a)]
is constructed from a rhombus with an acute angle of $\frac{\pi}{6}$, a square, and
a hexagon [Fig.~\ref{F:SocolarLattice}(b)].
The Socolar lattice is obtained as a projection of a six-dimensional hypercubic lattice
onto a two-dimensional real space,
\begin{align}
   &
   \left[
   \begin{array}{c}
     x
    \\
     y
    \\
     X
    \\
     Y
    \\
     i
    \\
     j
   \end{array}
   \right]
  =
   \frac{1}{\sqrt{3}}
   \left[
   \begin{array}{cccccc}
     c_{0} & c_{1} & -c_{2} & -c_{3} & c_{4} & c_{5}
    \\
     s_{0} & s_{1} & -s_{2} & -s_{3} & s_{4} & s_{5}
    \\
     c_{0} & c_{5} & c_{4} & -c_{3} & -c_{2} & c_{1}
    \\
     s_{0} & s_{5} & s_{4} & -s_{3} & -s_{2} & s_{1}
    \\
     0 & 1 & 0 & 1 & 0 & 1
    \\
     1 & 0 & 1 & 0 & 1 & 0
   \end{array}
   \right]
   \left[
   \begin{array}{c}
     m_{1}
    \\
     m_{2}
    \\
     m_{3}
    \\
     m_{4}
    \\
     m_{5}
    \\
     m_{6}
   \end{array}
   \right];
   \allowdisplaybreaks
   \nonumber \\
   &
   c_{n}=\cos\left(\frac{\pi}{6}n\right),
  \ 
   s_{n}=\sin\left(\frac{\pi}{6}n\right),
  \ 
   m_{n}\in\mathbb{Z},
\label{E:projectionSocolar}
\end{align}
where $(x,y)$ and $(X,Y,i,j)$ are the coordinates in the two-dimensional real and
four-dimensional perpendicular spaces, respectively.
The prefactor $\sqrt{1/3}$ serves as the lattice constant.
Every vertex of the Socolar lattice is represented by an integer linear combination of four
independent primitive lattice vectors [Fig.~\ref{F:SocolarLattice}(b)].
The infinite Socolar lattice consists of five types of local vertices, whose coordination
numbers $z$ range from 3 to 6 [Fig.~\ref{F:SocolarLattice}(c)].
The projection windows in the four-dimensional perpendicular space of the Socolar lattice
are hexagons lying in the subspaces spanned by $(\sqrt{3}i,\sqrt{3}j)=(1,1)$, $(1,2)$,
$(2,1)$, and $(2,2)$ [Fig.~\ref{F:SocolarLattice}(d)].
The vertex sites contained in the $(1,1)$ and $(2,2)$ planes form the even sublattice,
whereas those in the $(1,2)$ and $(2,1)$ planes form the odd sublattice.
The areas and structures of the hexagonal projection windows of the Socolar lattice do
not depend on planes, unlike the two distinct pentagonal projection windows of the Penrose
lattice.
Thus, a single representative plane suffices for perpendicular-space analysis.
The infinite Socolar lattice has dodecagonal symmetry, with the LI class invariant under
corresponding symmetry operations.

\section{Model and Method}
   We study the spin $S=\frac{1}{2}$ nearest-neighbor antiferromagnetic \textit{XXZ} model
on the two-dimensional $\mathbf{C}_{5\mathrm{v}}$ Penrose, $\mathbf{C}_{8\mathrm{v}}$
Ammann-Beenker, and $\mathbf{C}_{3\mathrm{v}}$ Socolar tilings, each consisting of bipartite
sublattices A with $L_{\mathrm{A}}$ sites and B with $L_{\mathrm{B}}$ sites
($L_{\mathrm{A}}+L_{\mathrm{B}}\equiv L$), whose Hamiltonian reads
\begin{align}
   \mathcal{H}
  =J\sum_{\langle i,j \rangle}
   \left\{
    \varDelta\left( S_{\bm{r}_{i}}^{x}S_{\bm{r}_{j}}^{x}
                        +S_{\bm{r}_{i}}^{y}S_{\bm{r}_{j}}^{y}\right)
   +S_{\bm{r}_{i}}^{z}S_{\bm{r}_{j}}^{z}
   \right\},
\label{E:XXZHamiltonian}
\end{align}
where $0<J$ and $0\leq\varDelta\leq 1$.
We denote the vector spin operators on the A and B sublattices by $\bm{S}_{\bm{r}_{i}}$
($i=1,\cdots,L_{\mathrm{A}}$) and $\bm{S}_{\bm{r}_{j}}$ ($j=1,\cdots,L_{\mathrm{B}}$),
respectively, with a tacit understanding of denoting an arbitrary---in the sense of possibly
running over both sublattices---site by $\bm{r}_{l}$.
The summation over all and only nearest neighbors $\sum_{\langle i,j \rangle}$ can be calculated as
$
   \sum_{i\in\mathrm{A}}\sum_{j\in\mathrm{B}}
   l_{i,j}
$
with the linkage identifier
$l_{i,j}$ being $1$ for connected vertices $\bm{r}_{i}$ and $\bm{r}_{j}$, otherwise $0$.
We express the Hamiltonian \eqref{E:XXZHamiltonian} in terms of the Holstein-Primakoff
bosons \cite{H1098},
\begin{align}
   &
   S_{\bm{r}_{i}}^{+}
  =\left(2S-a_{i}^{\dagger}a_{i}\right)^{\frac{1}{2}}
   a_{i},\ \ 
   S_{\bm{r}_{j}}^{+}
  =b_{j}^{\dagger}
   \left(2S-b_{j}^{\dagger}b_{j}\right)^{\frac{1}{2}},
   \allowdisplaybreaks
   \nonumber \\
   &
   S_{\bm{r}_{i}}^{-}
  =a_{i}^{\dagger}
   \left(2S-a_{i}^{\dagger}a_{i}\right)^{\frac{1}{2}},\ 
   S_{\bm{r}_{j}}^{-}
  =\left(2S-b_{j}^{\dagger}b_{j}\right)^{\frac{1}{2}}
   b_{j},
   \allowdisplaybreaks
   \nonumber \\
   &
   S_{\bm{r}_{i}}^{z}
  =S-a_{i}^{\dagger}a_{i},\qquad\quad\ 
   S_{\bm{r}_{j}}^{z}
  =b_{j}^{\dagger}b_{j}-S.
   \label{E:HPboson}
\end{align}
We expand \eqref{E:XXZHamiltonian} in descending powers of  $S$ \cite{Y11033},
\begin{align}
   &
    \mathcal{H}
   =\mathcal{H}^{(2)}
   +\mathcal{H}^{(1)}
   +O(S^{0});
   \allowdisplaybreaks
   \nonumber \\
   &
   \mathcal{H}^{(2)}
 =-JS^2
   \sum_{i\in\mathrm{A}}\sum_{j\in\mathrm{B}}
   l_{i,j},
   \allowdisplaybreaks
   \nonumber \\
   &
   \mathcal{H}^{(1)}
  =JS
   \sum_{i\in\mathrm{A}}\sum_{j\in\mathrm{B}}
   l_{i,j}
   \left\{
    a_{i}^{\dagger}a_{i}
   +b_{j}^{\dagger}b_{j}
   +\varDelta\left( a_{i}b_{j}
                   +a_{i}^{\dagger}b_{j}^{\dagger}\right)
   \right\}.
\label{E:H2&H1}
\end{align}
We define the up-to-$O(S^{1})$ linear SW (LSW) Hamiltonian
\begin{align}
   \mathcal{H}_{\mathrm{LSW}}
  \equiv
   \mathcal{H}^{(2)}
  +\mathcal{H}^{(1)}.
   \label{E:LSWHam}
\end{align}

   We introduce the row vectors $\bm{a}^{\dagger}$, ${}^{t}\bm{b}$, and $\bm{c}^{\dagger}$,
of dimension $L_{\mathrm{A}}$, $L_{\mathrm{B}}$, and $L$, respectively, as
\begin{align}
   \bm{c}^{\dagger}
  =\left[a_{1}^{\dagger}, \cdots, a_{L_{\mathrm{A}}}^{\dagger},
         b_{1}, \cdots, b_{L_{\mathrm{B}}}\right]
  \equiv
   \left[\bm{a}^{\dagger}, {}^{t}\bm{b}\right].
\label{E:HProwvector}
\end{align}
Then, the $O(S^{1})$ SW Hamiltonian can be expressed compactly in matrix form as
\begin{align}
   &
   \mathcal{H}^{(1)}
  =\varepsilon^{(1)}
  +J
   \left[\bm{a}^{\dagger}, {}^{t}\bm{b}\right]
   \left[
   \begin{array}{c|c}
     \mathbf{Z}_{\mathrm{AA}}^{(1)} & \mathbf{C}_{\mathrm{AB}}^{(1)}
    \\ \hline
     \mathbf{C}_{\mathrm{AB}}^{(1)\dagger} & \mathbf{Z}_{\mathrm{BB}}^{(1)}
   \end{array}
   \right]
   \left[
   \begin{array}{c}
     \bm{a}
    \\
     {}^{t}\bm{b}^{\dagger}
   \end{array}
   \right]
   \allowdisplaybreaks
   \nonumber \\
   &
  \equiv
   \varepsilon^{(1)}
  +J\bm{c}^{\dagger}\mathbf{M}^{(1)}\bm{c};\ 
   \varepsilon^{(1)}
  =-JS\sum_{i\in\mathrm{A}}\sum_{j\in\mathrm{B}}l_{i,j},
\label{E:LSWHammatrix}
\end{align}
where $\mathbf{Z}_{\mathrm{AA}}^{(1)}$ and $\mathbf{Z}_{\mathrm{BB}}^{(1)}$ are diagonal
matrices of dimension $L_{\mathrm{A}}\times L_{\mathrm{A}}$ and
$L_{\mathrm{B}}\times L_{\mathrm{B}}$, respectively, while
$\mathbf{C}_{\mathrm{AB}}^{(1)}$ is a biadjacency matrix of dimension
$L_{\mathrm{A}}\times L_{\mathrm{B}}$.
The elements of these matrices are defined as
\begin{align}
   \left[\mathbf{Z}_{\mathrm{AA}}^{(1)}\right]_{l,l'}
  =\left[\mathbf{Z}_{\mathrm{BB}}^{(1)}\right]_{l,l'}
  =\delta_{l,l'}z_{l}S,
   \ 
   \left[\mathbf{C}_{\mathrm{AB}}^{(1)}\right]_{i,j}
  =l_{i,j}S\varDelta,
\label{E:blockdef}
\end{align}
where $z_{l}$ is the coordination number of the site at $\bm{r}_{l}$.

   We perform the  Bogoliubov transformation \cite{WA450,C327}
\begin{align}
   &
   \bm{c}
  =\left[
   \begin{array}{c|c}
     \mathbf{S} & \mathbf{U}
    \\ \hline
     \mathbf{V} & \mathbf{T}
   \end{array}
   \right]
   \bm{\alpha};\ 
   [\mathbf{S}]_{i,k_{-}}=s_{i,k_{-}},\ 
   [\mathbf{T}]_{j,k_{+}}=t_{j,k_{+}},
   \allowdisplaybreaks
   \nonumber \\
   &
   [\mathbf{U}]_{i,k_{+}}=u_{i,k_{+}},\ 
   [\mathbf{V}]_{j,k_{-}}=v_{j,k_{-}}
\label{E:Bogoliubov}
\end{align}
with the matrices $\mathbf{S}$, $\mathbf{T}$, $\mathbf{U}$, and $\mathbf{V}$,
of dimension 
$L_{\mathrm{A}}\times L_{-}$, 
$L_{\mathrm{B}}\times L_{+}$, 
$L_{\mathrm{A}}\times L_{+}$, and
$L_{\mathrm{B}}\times L_{-}$, respectively, to obtain quasiparticle magnons
\begin{align}
   \left[\alpha_{1}^{-\dagger},\cdots,\alpha_{L_{-}}^{-\dagger},
         \alpha_{1}^{+},\cdots,\alpha_{L_{+}}^{+}\right]
  \equiv
   \bm{\alpha}^{\dagger}
\label{E:QPmagnonvector}
\end{align}
and their Hamiltonian
\begin{align}
   \mathcal{H}_{\mathrm{LSW}}
  =\mathcal{H}^{(2)}
  +\varepsilon^{(1)}
  +\sum_{k_{+}=1}^{L_{+}}\varepsilon_{k_{+}}^{+}
  +\sum_{\sigma=\mp}\sum_{k_{\sigma}=1}^{L_{\sigma}}
   \varepsilon_{k_{\sigma}}^{\sigma}
   \alpha_{k_{\sigma}}^{\sigma\dagger}\alpha_{k_{\sigma}}^{\sigma},
\label{E:LSWdiagHam}
\end{align}
where $\alpha_{k_{\sigma}}^{\sigma\dagger}$ creates a ferromagnetic ($\sigma=-$) or
antiferromagnetic ($\sigma=+$) magnon of energy $\varepsilon_{k_{\sigma}}^{\sigma}$,
which reduces or enhances the ground-state magnetization \cite{YR14008}, respectively.
In the thermodynamic limit $L\to\infty$, the A and B sublattices are equivalent,
making the system symmetric under a global reversal of the magnetization axis and the
$\sigma=-$ and $\sigma=+$ modes equivalent.

\section{First-Order Spin-Orbit Raman Scattering}
   Single-magnon-mediated Raman scattering is described as a third-order transition,
involving a second-order electric-dipole coupling and a first-order SO interaction
\cite{F514}.
The matrix element for the Raman scattering process at the magnetic ion on site $\bm{r}_{l}$
is given by
\begin{align}
   &
   M_{l}
  =\sum_{m,n=-1}^{1}
   \frac{\hbar\sqrt{\omega_{\mathrm{in}}\omega_{\mathrm{sc}}}e^{2}\lambda}{2\varepsilon_{0}V}
   \left[ \frac{1}{(E_{0}-\hbar\omega_{\mathrm{in}})^{2}}
         -\frac{1}{(E_{0}+\hbar\omega_{\mathrm{sc}})^{2}} \right]
   \allowdisplaybreaks
   \nonumber \\
   &
   \times
   \langle S,0|\bm{e}_{\mathrm{sc}}^{*}\cdot\bm{d}|P,n \rangle
   \langle P,n|\bm{L}_{\bm{r}_{l}}\cdot\bm{S}_{\bm{r}_{l}}|P,m \rangle
   \langle P,m|\bm{e}_{\mathrm{in}}\cdot\bm{d}|S,0 \rangle.
\label{E:SOtransitionmatrix}
\end{align}
We denote the elementary charge by $e$ and the electric dipole moment by $-e\bm{d}$.
$\lambda$ is the SO coupling constant, and $\bm{L}_{\bm{r}_{l}}$ represents the
orbital angular momentum operator.
$\bm{e}_{\mathrm{in}}$ and $\bm{e}_{\mathrm{sc}}$ denote the polarization vectors of the
incident and scattered photons, respectively, with the $z$ axis taken along the spin easy
axis.
$\omega_{\mathrm{in}}$ and $\omega_{\mathrm{sc}}$ denote the frequencies of the incident
and scattered photons, $\varepsilon_{0}$ is the vacuum permittivity, and $V$ is the system
volume.
In the bra-ket notation, capital letters indicate the orbital angular momentum quantum
number ($S$ and $P$ correspond to 0 and 1, respectively), while integers indicate the value
of its $z$ component.
$E_{0}$ is the orbital excitation energy between the $S$ and $P$ orbitals, which is assumed
to be much larger than both $\lambda$ and the magnon excitation energies.

   The matrix element is summed over all sites and mapped onto the spin sector.
The effective operator for SO-mechanism magnetic Raman scattering
\cite{F514,K2993} is given by
\begin{align}
   &
   \mathcal{R}_{\mathrm{SO}}
  =i\Gamma\sum_{l=1}^{L}
   \left(\bm{e}_{\mathrm{in}}\times\bm{e}_{\mathrm{sc}}^{*}\right)
   \cdot\bm{S}_{\bm{r}_{l}};
   \allowdisplaybreaks
   \nonumber \\
   &
   \Gamma
  =\frac{\hbar\sqrt{\omega_{\mathrm{in}}\omega_{\mathrm{sc}}}e^{2}\lambda |\bm{d}|^{2}}
        {6\varepsilon_{0}V}
   \left[ \frac{1}{(E_{0}-\hbar\omega_{\mathrm{in}})^{2}}
         -\frac{1}{(E_{0}+\hbar\omega_{\mathrm{sc}})^{2}} \right].
\label{E:SORamanoperator}
\end{align}
This scattering is purely antisymmetric with respect to the incident polarization
$\bm{e}_{\mathrm{in}}$ and the scattered polarization $\bm{e}_{\mathrm{sc}}$, and it
vanishes for parallel $\bm{e}_{\mathrm{in}}$ and $\bm{e}_{\mathrm{sc}}$, regardless of
their direction.

   Putting
$
   \mathcal{R}_{\mathrm{SO}}(t)
  \equiv
   e^{i\mathcal{H}_{\mathrm{LSW}}t/\hbar}
   \mathcal{R}_{\mathrm{SO}}
   e^{-i\mathcal{H}_{\mathrm{LSW}}t/\hbar}
$,
the scattering intensity at absolute zero is obtained from
\begin{align}
   I(\omega)
  =\frac{1}{2\pi\hbar L}\int_{-\infty}^{\infty}
   e^{i\omega t}
   \left\langle
     \mathcal{R}_{\mathrm{SO}}^{\dagger}(t)\mathcal{R}_{\mathrm{SO}}(0)
   \right\rangle
   dt,
\label{E:RamanInt}
\end{align}
where $\langle \cdots \rangle$ denotes a quantum average in the magnon vacuum.
Expanding each component of \eqref{E:SORamanoperator} yields
\begin{align}
   \mathcal{R}_{\mathrm{SO}}
  =&
   \frac{1}{2}\Gamma\sum_{l=1}^{L}
   \left(
     P^{-}S_{\bm{r}_{l}}^{-}
    +P^{+}S_{\bm{r}_{l}}^{+}
    +P^{z}S_{\bm{r}_{l}}^{z}
   \right),
\label{E:SORamanoperatorEX}
\end{align}
where
$
   P^{-}
  =e_{\mathrm{in}}^{+}e_{\mathrm{sc}}^{*z}
  -e_{\mathrm{in}}^{z}e_{\mathrm{sc}}^{*+}
$, 
$
   P^{+}
  =e_{\mathrm{in}}^{z}e_{\mathrm{sc}}^{*-}
  -e_{\mathrm{in}}^{-}e_{\mathrm{sc}}^{*z}
$, 
$
   P^{z}
  =e_{\mathrm{in}}^{-}e_{\mathrm{sc}}^{*+}
  -e_{\mathrm{in}}^{+}e_{\mathrm{sc}}^{*-}
$, 
and
$
   e_{\mathrm{in}}^{\pm}
  =e_{\mathrm{in}}^{x}\pm ie_{\mathrm{in}}^{y}
$, 
$
   e_{\mathrm{sc}}^{*\pm}
  =e_{\mathrm{sc}}^{*x}\pm ie_{\mathrm{sc}}^{*y}
$.
Owing to the conservation of magnetization, the transverse components
($P^{-}S_{\bm{r}_{l}}^{-}$ and $P^{+}S_{\bm{r}_{l}}^{+}$) are orthogonal to the
longitudinal component ($P^{z}S_{\bm{r}_{l}}^{z}$).
We therefore concentrate on the single-magnon scattering due to the transverse components
by taking the scattered polarization vector to be the linear polarization along the
$z$-axis (the spin easy axis), $\bm{e}_{\mathrm{sc}}=(0,0,1)$.

   If we truncate the effective Raman operator \eqref{E:SORamanoperatorEX} at the order
of $S^{\frac{1}{2}}$, the transverse contribution to the SO-mechanism
first-order Raman scattering is given by
\begin{align}
   I^{\perp}(\omega)
  =\frac{\Gamma^{2}S}{2L}
   \sum_{\sigma=\mp}
   \left|P^{\sigma}\right|^{2}\sum_{k_{\sigma}=1}^{L_{\sigma}}
   \left|C_{\sigma}(k_{\sigma})\right|^{2}
   \delta\left(\hbar\omega-\varepsilon_{k_{\sigma}}^{\sigma}\right),
\label{E:RamanIntEq}
\end{align}
where the coefficients $C_{\sigma}(k_{\sigma})$ can be written in terms of the Bogoliubov
transformation matrix \eqref{E:Bogoliubov},
\begin{align}
   &
   C_{-}(k_{-})
  \equiv
   \sum_{l=1}^{L}\tilde{C}_{-}(l;k_{-})
  =\sum_{i\in\mathrm{A}}s_{i,k_{-}}
  +\sum_{j\in\mathrm{B}}v_{j,k_{-}},
   \allowdisplaybreaks
   \nonumber \\
   &
   C_{+}(k_{+})
  \equiv
   \sum_{l=1}^{L}\tilde{C}_{+}(l;k_{+})
  =\sum_{i\in\mathrm{A}}u_{i,k_{+}}
  +\sum_{j\in\mathrm{B}}t_{j,k_{+}}.
\label{E:Ramancoefficient}
\end{align}

   We make the best use of light polarization in selectively observing a particular magnon mode.
Equation \eqref{E:SORamanoperator} shows that the present SO-mechanism Raman scattering is
forbidden when the incident and scattered polarizations are parallel.
If the scattered light polarization is $\bm{e}_{\mathrm{sc}}=(0,0,1)$,
the incident light polarization vector must be in the $xy$-plane.
In the case of incident light being linearly polarized,
$\bm{e}_{\mathrm{in}}=(\cos\theta,\sin\theta,0)$,
one finds $|P^{-}|^{2}=|P^{+}|^{2}=1$, i.e.,
both $\sigma=\mp$ modes are Raman active.
Note that the scattering intensity is independent of the polarization angle $\theta$.
In the case of incident light being right-handed (left-handed) circularly polarized,
$\bm{e}_{\mathrm{in}}=(1,-i,0)/\sqrt{2}$ [$\bm{e}_{\mathrm{in}}=(1,i,0)/\sqrt{2}$],
one finds $|P^{-}|^{2}=2$ and $|P^{+}|^{2}=0$ ($|P^{-}|^{2}=0$ and $|P^{+}|^{2}=2$),
i.e., only magnons of the $\sigma=-$ ($\sigma=+$) mode are Raman active, respectively.
Considering that the $\sigma=\pm$ magnon modes degenerate in the thermodynamic limit,
we calculate SO-mechanism Raman spectra with
$\bm{e}_{\mathrm{in}}=(1,0,0)$ and $\bm{e}_{\mathrm{sc}}=(0,0,1)$.
It is instructive to compare this polarization dependence with an alternative single-magnon
process recently proposed for Mott insulators with strong SO coupling, such as
Kitaev materials \cite{Y144412}.
In these systems, a prominent single-magnon channel arises from ligand-mediated hopping
in addition to the conventional photon-assisted virtual exchange via direct hopping between
magnetic ions.
In exchange-mediated processes, the polarization dependence is expressed in terms of the
scalar products of the polarization vectors and  bond directions
\cite{I2000118,I053701}.
The scattering intensity therefore strongly depends on the lattice symmetries.
This lattice-dependent nature forms a striking contrast to the selection rules for the
SO-mechanism, which are independent of the background lattice geometry.

\begin{figure}[tb]
\centering
\includegraphics[width=0.95\linewidth]{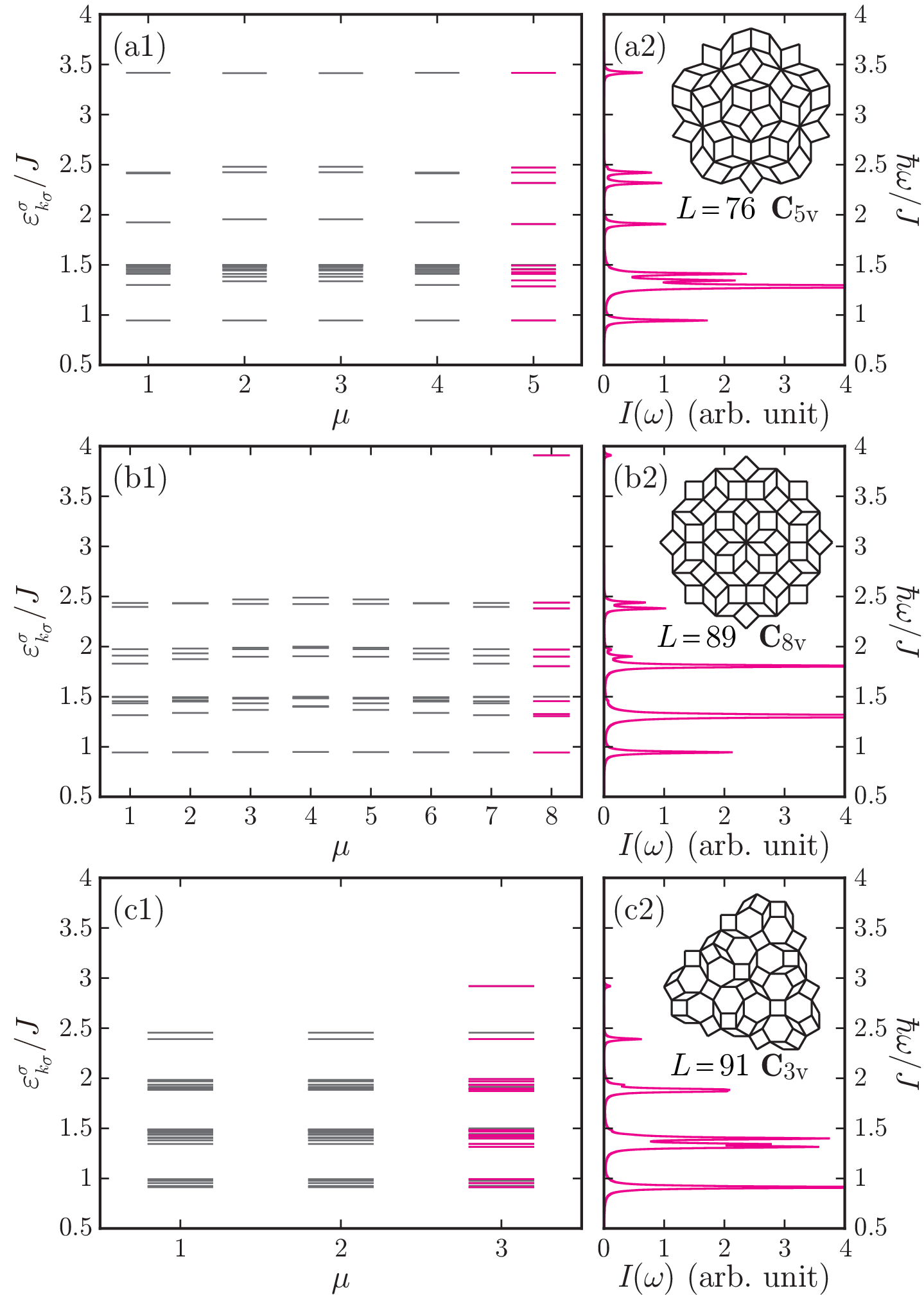}
\caption{%
         Eigenvalues $\varepsilon_{k_{\sigma}}^{\sigma}$ at $\varDelta=0.5$ of
         the $\mathbf{C}_{5\mathrm{v}}$ ($m=5$) Penrose cluster of $L=76$ (a1),
         the $\mathbf{C}_{8\mathrm{v}}$ ($m=8$) Ammann-Beenker cluster of $L=89$ (b1),
         and the $\mathbf{C}_{3\mathrm{v}}$ ($m=3$) Socolar cluster of $L=91$ (c1),
         resolved by the rotational quantum
         number $Q_{\mu}=2\pi\mu/m$ $(\mu=1,\cdots,m)$.
         Magenta levels indicate Raman-active states that are rotation-invariant
         ($Q_{\mu=m}=2\pi$) and mirror-symmetric ($P=2\pi$).
         Spin-orbit mechanism magnetic Raman spectra $I^{\perp}(\omega)$
         \eqref{E:RamanIntEq} for the Penrose (a2),
         Ammann-Beenker (b2), and Socolar (c2) clusters.
         Insets show schematic diagrams of the clusters.
         }
\label{F:levelspectrum}
\end{figure}

\subsection{Raman-Active Eigenstates}
   We first identify the eigenstates of Eq.~\eqref{E:LSWdiagHam} that are active in
SO-mechanism Raman scattering.
Any $\mathbf{C}_{m\mathrm{v}}$ cluster of $L$ sites is decomposable into $m$ fan-shaped rotational
units, each consisting of $(L-1)/m\equiv R$ sites, and the site of rotation center.
Introducing the rotational quantum number $Q_{\mu}=2\pi\mu/m$ ($\mu=1,\cdots,m$),
the matrix part of Hamiltonian \eqref{E:LSWHammatrix} reduces to an $(R+1)\times(R+1)$
matrix for $\mu=m$ and $R\times R$ matrices otherwise,
\begin{align}
   \mathcal{H}^{(1)}-\varepsilon^{(1)}
  =\bigoplus_{\mu=1}^{m}
   \mathcal{H}_{\mu}^{(1)}.
\label{E:rotationHamiltonian}
\end{align}
Each block Hamiltonian $\mathcal{H}_{\mu}^{(1)}$ has eigenvalue $\exp(iQ_{\mu})$ under
the $m$-fold rotation $C_{m}$.
Thus the magnon energy eigenvalues $\varepsilon_{k_{\sigma}}^{\sigma}$ can be classified
according to $\mu$.
For odd $m$, the $Q_{\mu=m}=2\pi$ sector, and for even $m$, both the $Q_{\mu=m}=2\pi$ and
$Q_{\mu=m/2}=\pi$ sectors correspond to one-dimensional irreducible representations.
The remaining sectors form degenerate pairs $(\mu, m-\mu)$, which correspond to
two-dimensional irreducible representations $\mathrm{E}_{\mu}$ ($\mu<m/2$).
The states with $Q_{\mu=m}=2\pi$ and $Q_{\mu=m/2}=\pi$ are simultaneously eigenstates of
the mirror operation $\sigma_{\mathrm{v}}$ and can be further classified by the
eigenvalue $\exp(iP)$ into mirror-symmetric ($P=2\pi$) and mirror-antisymmetric
($P=\pi$) states.

   Figure~\ref{F:levelspectrum} shows the eigenvalues and their corresponding SO-mechanism
single-magnon Raman spectra for small clusters of each quasiperiodic lattice at
$\varDelta=0.5$.
The eigenvalues are plotted according to the rotational quantum number
$Q_{\mu}=2\pi\mu/m$.
Among the $Q_{\mu=m}=2\pi$ levels, mirror-symmetric ones are highlighted in magenta.
Only rotation-invariant ($Q_{\mu=m}=2\pi$) and mirror-symmetric ($P=2\pi$) states are
Raman active in single-magnon scattering, corresponding to the $\mathrm{A}_{1}$
irreducible representation of the $\mathbf{C}_{m\mathrm{v}}$ point group.

   The selection rules for the Raman-active levels found here are entirely different from
those for the exchange-mechanism two-particle-mediated Raman scattering. 
For systems with point group $\mathbf{C}_{m\mathrm{v}}$ ($m \neq 2,4$), the lowest-order
nonvanishing Loudon-Fleury vertex \cite{F514} yields Raman-active modes belonging to the
two-dimensional irreducible representation $\mathrm{E}$ (for $m=3$) or $\mathrm{E}_{2}$
(for other $m$) \cite{I053701,K024414,K214411,Y063701}.
This condition requires that the sum of the rotational quantum numbers of the two
mediating particles satisfies $Q_{\mu_{1}}+Q_{\mu_{2}}=\pm 4\pi/m$.
While the exchange mechanism cannot select any particular magnon,
the SO mechanism uniquely selects the $\mathrm{A}_{1}$ species.

\begin{figure}[tb]
\centering
\includegraphics[width=\linewidth]{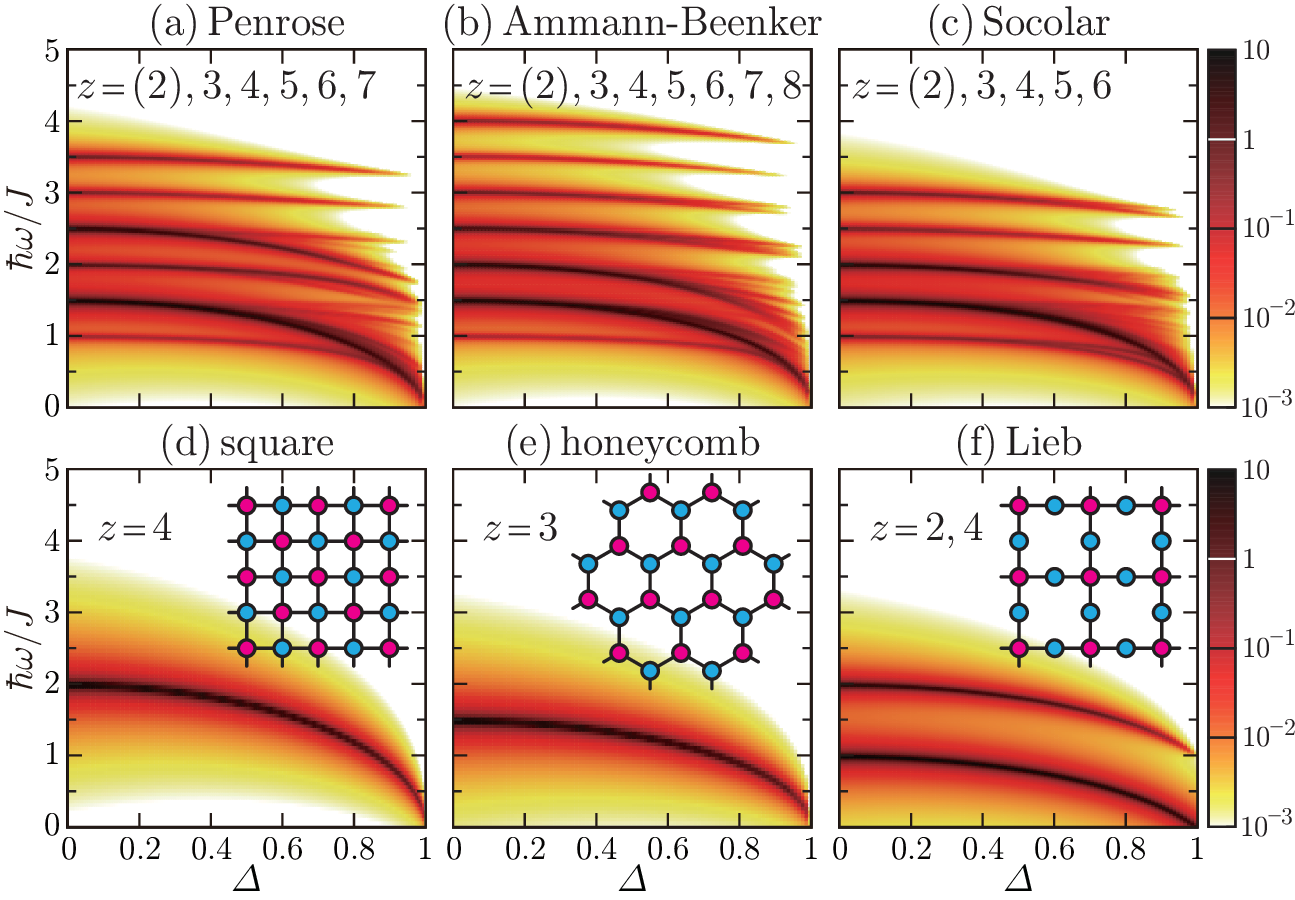}
\caption{%
         Contour plots of the spin-orbit mechanism magnetic Raman spectra
         $I^{\perp}(\omega)$ \eqref{E:RamanIntEq} as a function of the exchange anisotropy
         $\varDelta$ for bipartite quasiperiodic lattices with open boundaries and
         for periodic lattices.
         Penrose lattice of $L=10351$ (a), Ammann-Beenker lattice of $L=10457$ (b),
         Socolar lattice of $L=11566$ (c), square lattice (d), honeycomb lattice (e),
         and Lieb lattice (f).
         Note that $z=2$ vertices are artifacts of finite open-boundary clusters.
         Insets show schematic diagrams of the periodic lattices.
         }
\label{F:SORamandeltdepQP&P}
\end{figure}

\begin{figure}[tb]
\centering
\includegraphics[width=\linewidth]{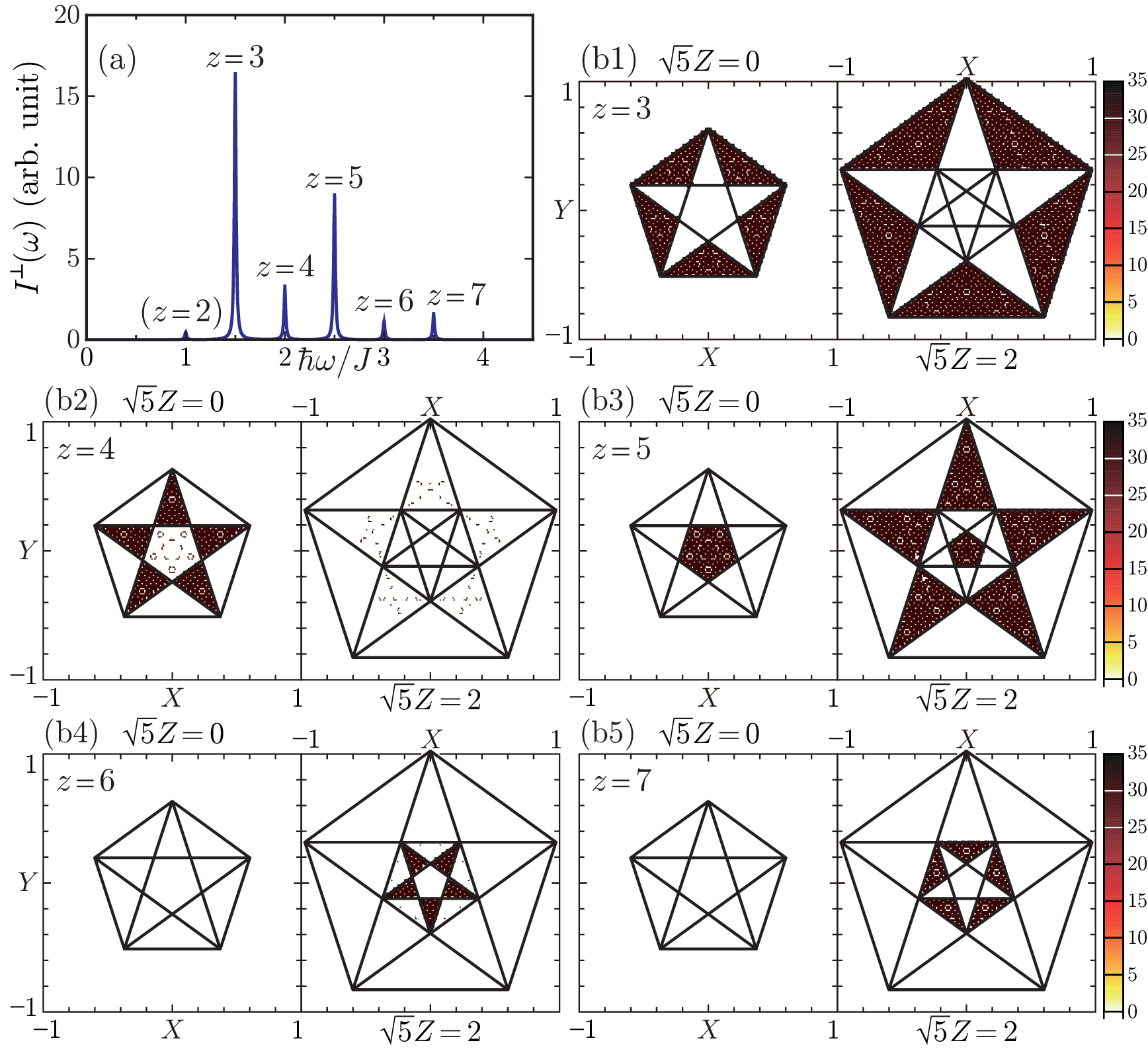}
\caption{%
         Spin-orbit-mechanism magnetic Raman spectrum $I^{\perp}(\omega)$
         \eqref{E:RamanIntEq} in the Ising limit ($\varDelta=0$) for the Penrose
         lattice of $L=10351$ (a).
         Contour plots of site-resolved Raman spectra $I^{\perp}(\omega)|_{l}$
         \eqref{E:siteresolvedRaman} multiplied by the number of sites in the perpendicular
         space at $\hbar\omega=zJS$ with $z=3$ to $7$ (b1-b5).
         The projection windows for $\sqrt{5}Z=0$ and $2$
         are shown.
         }
\label{F:perpPenroseIsing}
\end{figure}

\subsection{Anisotropy Dependence}
   Contour plots of $I^{\perp}(\omega)$ as functions of frequency $\hbar\omega$ and
exchange anisotropy $\varDelta$ for bipartite lattices---the quasiperiodic Penrose,
Ammann-Beenker, and
Socolar lattices, and the periodic square, honeycomb, and Lieb \cite{N063622,S041410R}
lattices---are shown in Fig.~\ref{F:SORamandeltdepQP&P}.
In the Ising limit ($\varDelta=0$), the Raman spectrum exhibits delta-function peaks at
degenerate energies $zJS$ determined by the coordination number $z$, whether the lattice
is periodic or quasiperiodic.
In the Heisenberg isotropic point ($\varDelta=1$), the Hamiltonian \eqref{E:XXZHamiltonian}
commutes with the effective Raman operator \eqref{E:SORamanoperator}, and thus the spectral
weight vanishes. 
In the intermediate regime ($0<\varDelta<1$), the lifting of the eigenvalue degeneracy shifts
the peaks toward lower energies.
Delta-function peaks bifurcate or divide into more in each individual manner on quasiperiodic
lattices, while it remains singly peaked all the way on periodic lattices.
Single-magnon Raman scattering in periodic systems is restricted to $\bm{k}=\bm{0}$ magnons
\cite{F514,K2993} by momentum conservation.
In contrast, quasiperiodic systems lack translational symmetry, breaking the wavevector
selection rule and allowing magnons with various momenta to contribute.
From a real-space perspective, the transverse exchange $\varDelta$ enables magnons to hop
between adjacent sites, making their excitation energies sensitive to local environments
beyond the nearest neighbors and thereby distinguishing even sites sharing the same $z$.
Such structural diversity lifts the eigenvalue degeneracy and leads to
the observed peak splittings.
As $\varDelta$ increases, the magnons explore a wider spatial range, further resolving
the energy differences between distinct local environments.

\begin{figure}[tb]
\centering
\includegraphics[width=\linewidth]{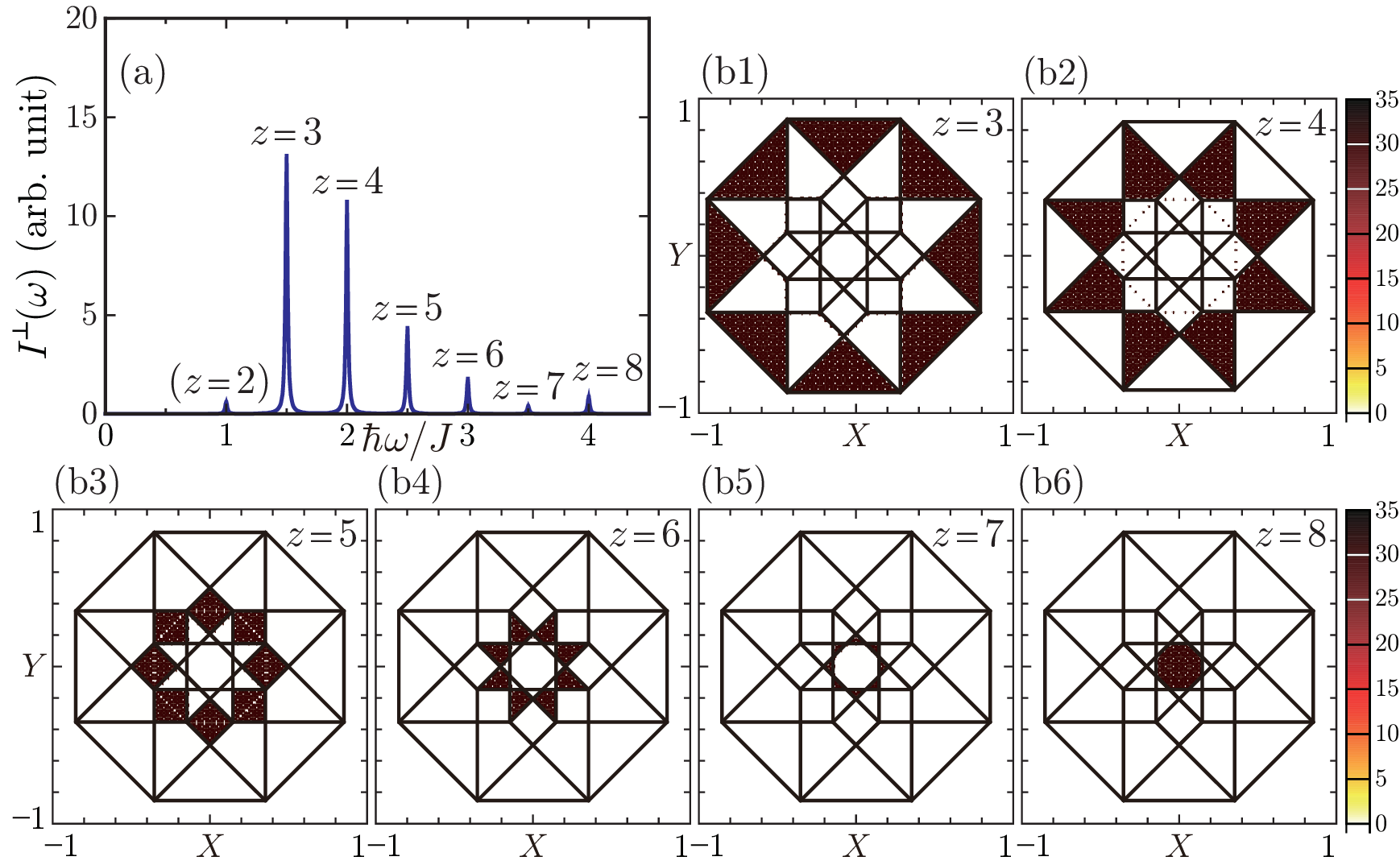}
\caption{%
         The same as Fig. \ref{F:perpPenroseIsing} for the Ammann-Beenker lattice of
         $L=10457$ and $z=3$ to $8$ [(a) and (b1-b6)].
         }
\label{F:perpABIsing}
\end{figure}
\begin{figure}[tb]
\centering
\includegraphics[width=0.75\linewidth]{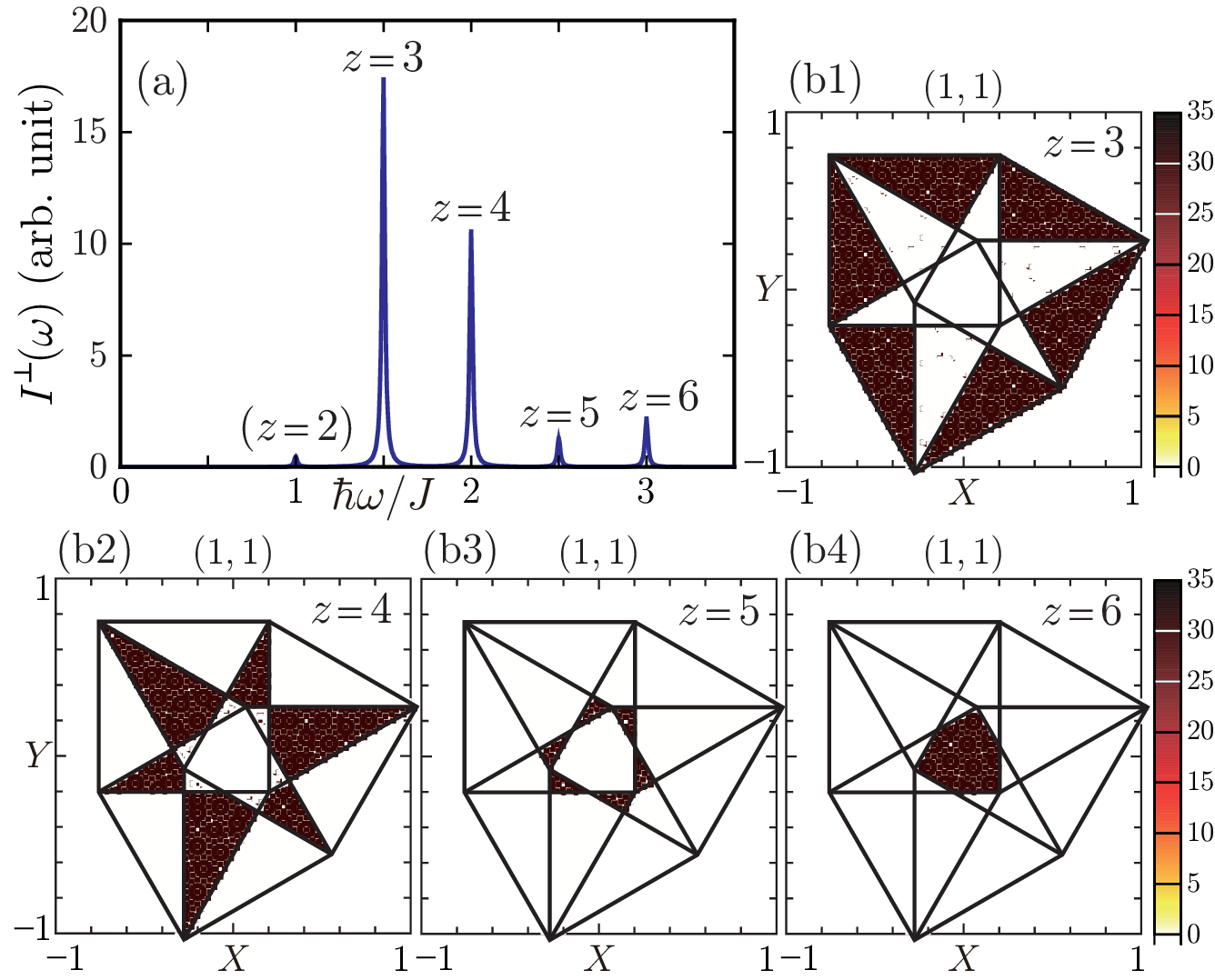}
\caption{%
         The same as Fig. \ref{F:perpPenroseIsing} for the Socolar lattice of $L=11566$
         and $z=3$ to $6$ [(a) and (b1-b4)].
         The projection window for
         $(\sqrt{3}i,\sqrt{3}j)=(1,1)$ is shown.
         }
\label{F:perpSocolarIsing}
\end{figure}

\begin{figure*}[tb]
\centering
\includegraphics[width=0.95\linewidth]{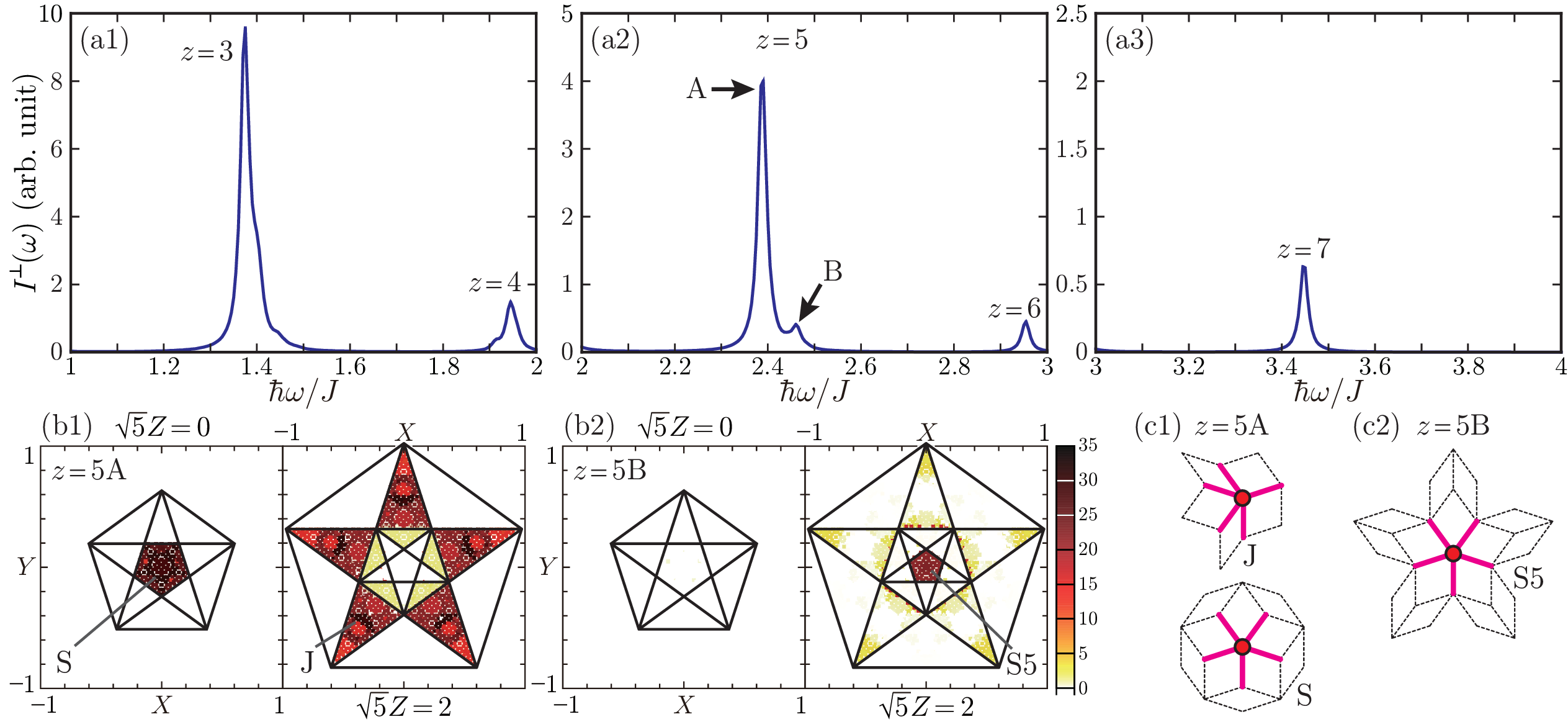}
\caption{%
         Spin-orbit-mechanism magnetic Raman spectra $I^{\perp}(\omega)$
         \eqref{E:RamanIntEq} for the Penrose lattice of $L=10351$ at
         $\varDelta=0.4$, shown for the frequency ranges
         $J \leq \hbar\omega \leq 2J$ (a1), $2J \leq \hbar\omega \leq 3J$ (a2), and
         $3J \leq \hbar\omega \leq 4J$ (a3).
         Contour plots of the site-resolved Raman spectra $I^{\perp}(\omega)|_{l}$
         \eqref{E:siteresolvedRaman} multiplied by the number of sites in the perpendicular
         space at the two split peaks indicated by arrows in (a2), associated with sites
         of coordination number $z=5$ (b1, b2).
         Real-space local environments corresponding to each branch of the $z=5$ peak
         (c1, c2).
         }
\label{F:perpPenrose0.4}
\end{figure*}
\begin{figure*}[tb]
\centering
\includegraphics[width=0.95\linewidth]{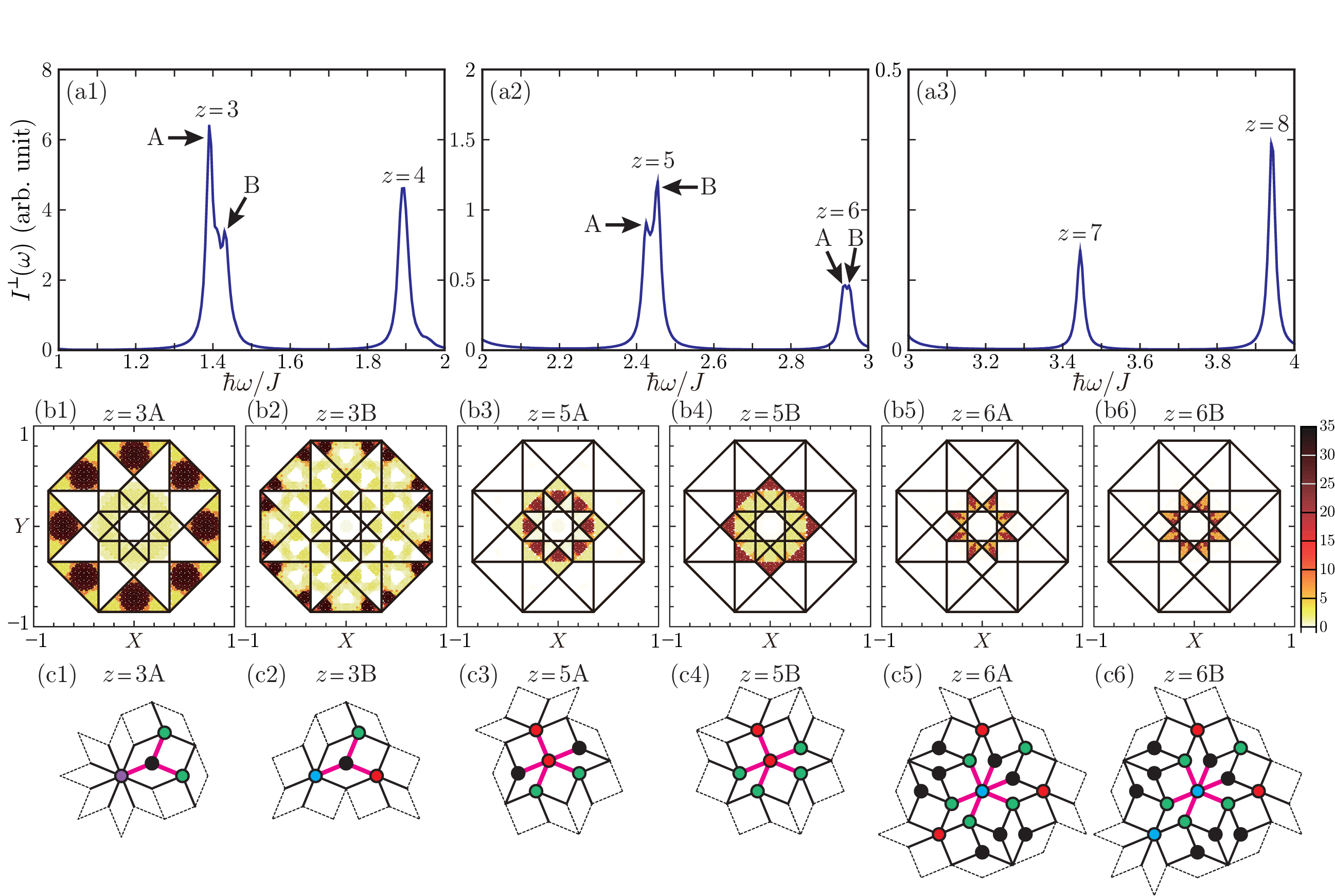}
\caption{%
         Spin-orbit-mechanism magnetic Raman spectra $I^{\perp}(\omega)$
         \eqref{E:RamanIntEq} for the Ammann-Beenker lattice of $L=10457$ at
         $\varDelta=0.4$, shown for the frequency ranges
         $J \leq \hbar\omega \leq 2J$ (a1), $2J \leq \hbar\omega \leq 3J$ (a2), and
         $3J \leq \hbar\omega \leq 4J$ (a3).
         Contour plots of site-resolved Raman spectra $I^{\perp}(\omega)|_{l}$
         \eqref{E:siteresolvedRaman} multiplied by the number of sites in the perpendicular
         space at the characteristic peaks, indicated by arrows in (a1) and (a2), associated
         with the sites of coordination numbers $z=3$, $z=5$, and $z=6$ (b1-b6).
         Corresponding real-space local environments for each peak (c1-c6).
         The site coloring follows Fig.~\ref{F:ABLattice}(c).
         }
\label{F:perpAB0.4}
\end{figure*}
\begin{figure*}[tb]
\centering
\includegraphics[width=0.95\linewidth]{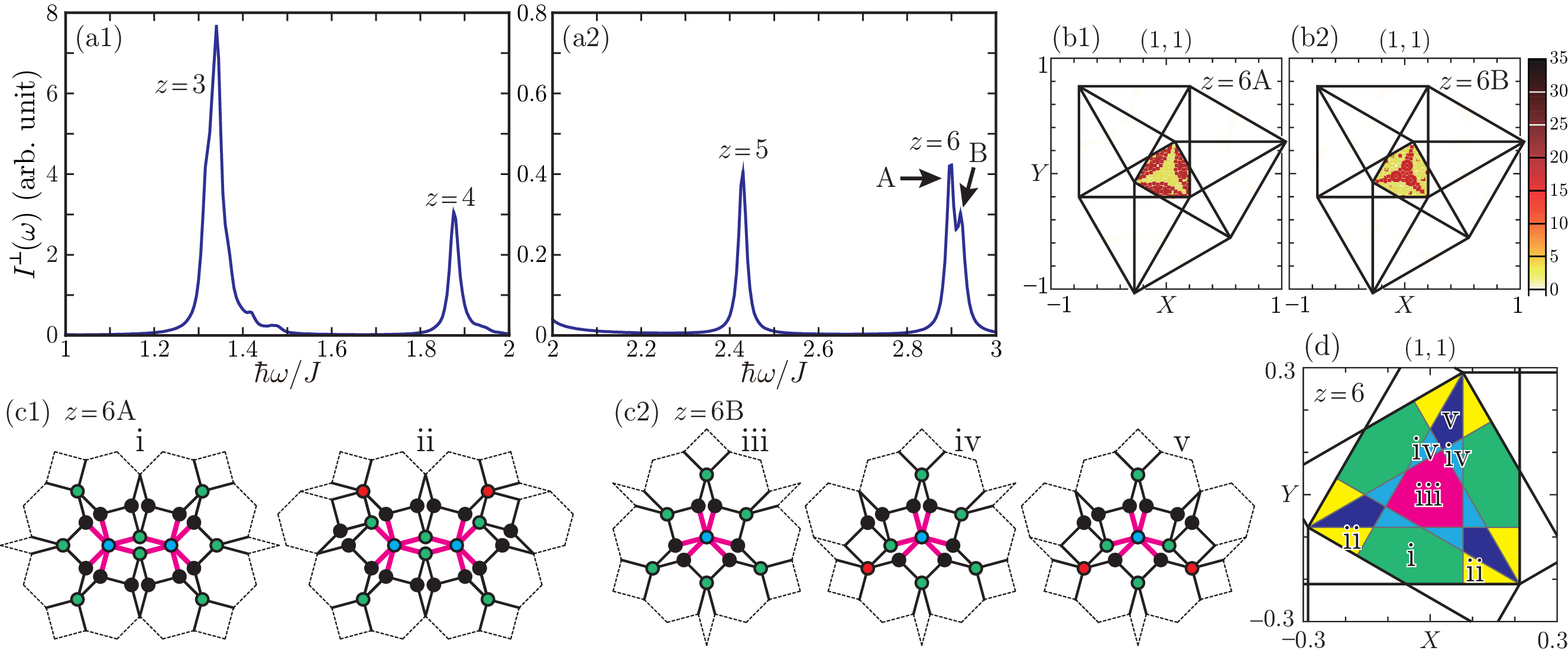}
\caption{%
         Spin-orbit-mechanism magnetic Raman spectra $I^{\perp}(\omega)$
         \eqref{E:RamanIntEq} for the Socolar lattice of $L=11566$ at
         $\varDelta=0.5$, shown for the frequency ranges
         $J \leq \hbar\omega \leq 2J$ (a1) and $2J \leq \hbar\omega \leq 3J$ (a2).
         Contour plots of the site-resolved Raman spectra $I^{\perp}(\omega)|_{l}$
         \eqref{E:siteresolvedRaman} multiplied by the number of sites in the perpendicular
         space at the two peaks, indicated by arrows in (a2), both associated with
         coordination number $z=6$ (b1, b2).
         Real-space local environments of $z=6$ sites categorized into five subtypes (i-v)
         defined by their next-nearest-neighbor configurations (c1, c2).
         The site coloring follows Fig.~\ref{F:SocolarLattice}(c).
         Partitioning of the $z=6$ domain in the perpendicular space into five
         subdomains (labeled i-v) corresponding to the subtypes in (c1) and (c2), where
         regions of the same color represent the same subtype (d).
         }
\label{F:perpSocolar0.5}
\end{figure*}

\subsection{Perpendicular-Space Analysis}
   To further evaluate the contribution of each site to the scattering intensity,
we define the site-resolved Raman spectrum
\begin{align}
   &
   I^{\perp}(\omega)
  \equiv
   \sum_{l=1}^{L}I^{\perp}(\omega)|_{l}
  =\sum_{l=1}^{L}\frac{\Gamma^{2}S}{2L}
   \sum_{\sigma=\mp}
   \left|P^{\sigma}\right|^{2}
   \allowdisplaybreaks
   \nonumber \\
   &\times
   \sum_{k_{\sigma}=1}^{L_{\sigma}}
   \mathrm{Re}\left[
     \tilde{C}_{\sigma}(l;k_{\sigma})C_{\sigma}(k_{\sigma})^{*}
   \right]
   \delta\left(\hbar\omega-\varepsilon_{k_{\sigma}}^{\sigma}\right).
\label{E:siteresolvedRaman}
\end{align}

   First, we discuss the Ising limit ($\varDelta=0$).
Figures~\ref{F:perpPenroseIsing}-\ref{F:perpSocolarIsing} show
the results for the Penrose, Ammann-Beenker, and Socolar lattices, respectively.
In each case, the Raman spectrum $I^{\perp}(\omega)$ [Figs.~\ref{F:perpPenroseIsing}(a),
\ref{F:perpABIsing}(a), and \ref{F:perpSocolarIsing}(a)] exhibits delta-function peaks
at $\hbar\omega=zJS$.
Contour plots of the site-resolved Raman spectra $I^{\perp}(\omega)|_{l}$ in the
perpendicular space at these peak frequencies are shown in
Figs.~\ref{F:perpPenroseIsing}(b1-b5), \ref{F:perpABIsing}(b1-b6), and 
\ref{F:perpSocolarIsing}(b1-b4).
A comparison with the mapping between real-space local environments and perpendicular-space
projection windows [Figs.~\ref{F:PenroseLattice}(d), \ref{F:ABLattice}(d), and 
\ref{F:SocolarLattice}(d)] demonstrates that each peak at $\hbar\omega=zJS$ corresponds
uniquely to sites of coordination number $z$.
The contribution from each site  is uniform across all peaks.
Thus the relative intensity  at $\hbar\omega=zJS$ is directly proportional to
the fraction of sites with coordination number $z$.

   Next, we consider the regime $0<\varDelta<1$.
The splittings of the Raman spectra emerge at approximately $\varDelta=0.4$ for the
Penrose and Ammann-Beenker lattices, and at $\varDelta=0.5$ for the Socolar lattice.
Figure \ref{F:perpPenrose0.4} presents the results for the Penrose lattice at
$\varDelta=0.4$.
The Raman spectra are shown for three frequency ranges:
$J \leq \hbar\omega \leq 2J$ [Fig.~\ref{F:perpPenrose0.4}(a1)],
$2J \leq \hbar\omega \leq 3J$ [Fig.~\ref{F:perpPenrose0.4}(a2)], and 
$3J \leq \hbar\omega \leq 4J$ [Fig.~\ref{F:perpPenrose0.4}(a3)].
The anisotropy $\varDelta$ lifts the eigenvalue degeneracy, shifting the peak positions
below $zJS$.
The peaks remain distinguishable by the coordination number $z$.
With increasing $\varDelta$, the linewidth broadens and the peak for $z=5$ splits into two.
The perpendicular-space representations of the site-resolved Raman spectra at
$\hbar\omega=2.39J$ and $\hbar\omega=2.46J$ are shown in
Figs.~\ref{F:perpPenrose0.4}(b1) and \ref{F:perpPenrose0.4}(b2), respectively.
Based on the domain classification of the projection windows [Fig.~\ref{F:PenroseLattice}(d)],
the lower branch is associated with the J- and S-type sites [Fig.~\ref{F:perpPenrose0.4}(c1)],
whereas the upper branch is associated with the S5-type sites [Fig.~\ref{F:perpPenrose0.4}(c2)].
The J- and S-type sites are  adjacent to at least two D-type sites ($z=3$), while
each S5-type site is surrounded by five J-type sites ($z=5$) \cite{Y702}.
This correspondence shows that the $z=5$ localized modes are stabilized when the coordination
numbers of adjacent sites markedly differ from that of the central site.

   The results for the Ammann-Beenker lattice at $\varDelta=0.4$ are shown in
Fig.~\ref{F:perpAB0.4}.
Distinct splittings are observed for $z=3$, $z=5$, and $z=6$ peaks
[Figs.~\ref{F:perpAB0.4}(a1-a3)].
Mapping the site-resolved Raman spectra for each peak onto the perpendicular space reveals
that the spectral weights are confined to specific subdomains [Figs.~\ref{F:perpAB0.4}(b1-b6)].
By partitioning the domains in the projection window based on the local environments including
more distant surroundings \cite{G224201,Y702}, we find a direct one-to-one correspondence
between the split peaks and the geometric subtypes.
For the $z=3$ and $z=5$ peaks [Figs.~\ref{F:perpAB0.4}(b1-b4)], this partitioning is fully
determined by the subclassification based on nearest-neighbor vertex configurations.
For the $z=6$ peaks [Figs.~\ref{F:perpAB0.4}(b5) and \ref{F:perpAB0.4}(b6)], the subdomains
are defined by their next-nearest-neighbor vertices.
The corresponding real-space environments are shown in Figs.~\ref{F:perpAB0.4}(c1-c6).
Consistent with  the Penrose lattice, the localized modes are stabilized
when the coordination numbers of surrounding sites deviate from that of the central site.
For instance, for the $z=3$ peaks, the mode adjacent to a $z=8$ site [Fig.~\ref{F:perpAB0.4}(c1)]
lies below that adjacent to a $z=6$ site [Fig.~\ref{F:perpAB0.4}(c2)].
Similarly, for the $z=5$ peaks, the mode adjacent to $z=3$ sites [Fig.~\ref{F:perpAB0.4}(c3)]
is lower than that adjacent to $z=4$ sites [Fig.~\ref{F:perpAB0.4}(c4)].
The one-to-one mapping demonstrates that local structural diversity lifts the eigenvalue
degeneracy.
Furthermore, the Ammann-Beenker lattice exhibits peak splittings over a broader range of
coordination numbers and energies than the Penrose lattice, underscoring the sensitivity of
excitations to the background lattice geometry in the presence of exchange anisotropy.

   Figure~\ref{F:perpSocolar0.5} shows the results for the Socolar lattice at $\varDelta=0.5$.
The peaks associated with $z=6$ sites exhibit a subtle splitting at $\hbar\omega=2.90J$ and
$2.92J$ [Fig.~\ref{F:perpSocolar0.5}(a2)].
The perpendicular-space representations of the site-resolved Raman spectra for each branch
are shown in Figs.~\ref{F:perpSocolar0.5}(b1) and \ref{F:perpSocolar0.5}(b2).
The spectral-weight distribution indicates that the $z=6$ domain consists of two major
regions, each associated with one of the split peaks.
We categorize the $z=6$ sites into five subtypes, (i)-(v), based on their local environments
up to the next-nearest neighbors
[Figs.~\ref{F:perpSocolar0.5}(c1) and \ref{F:perpSocolar0.5}(c2)]. 
The $z=6$ acceptance domain in the perpendicular space is partitioned into five corresponding
subdomains [Fig.~\ref{F:perpSocolar0.5}(d)].
Comparison of the spectral weights and the domain partitioning reveals that the lower branch
[Fig.~\ref{F:perpSocolar0.5}(b1)] originates from subdomains (i) and (ii), where  $z=6$
sites of the same subtype form next-nearest-neighbor pairs [Fig.~\ref{F:perpSocolar0.5}(c1)].
In contrast, the upper branch [Fig.~\ref{F:perpSocolar0.5}(b2)] corresponds to subdomains
(iii)-(v), which represent isolated $z=6$ sites [Fig.~\ref{F:perpSocolar0.5}(c2)].
These results indicate that the pairing of $z=6$ sites stabilizes the localized magnons
through the constructive overlap of their modes.

\begin{figure}[tb]
\centering
\includegraphics[width=\linewidth]{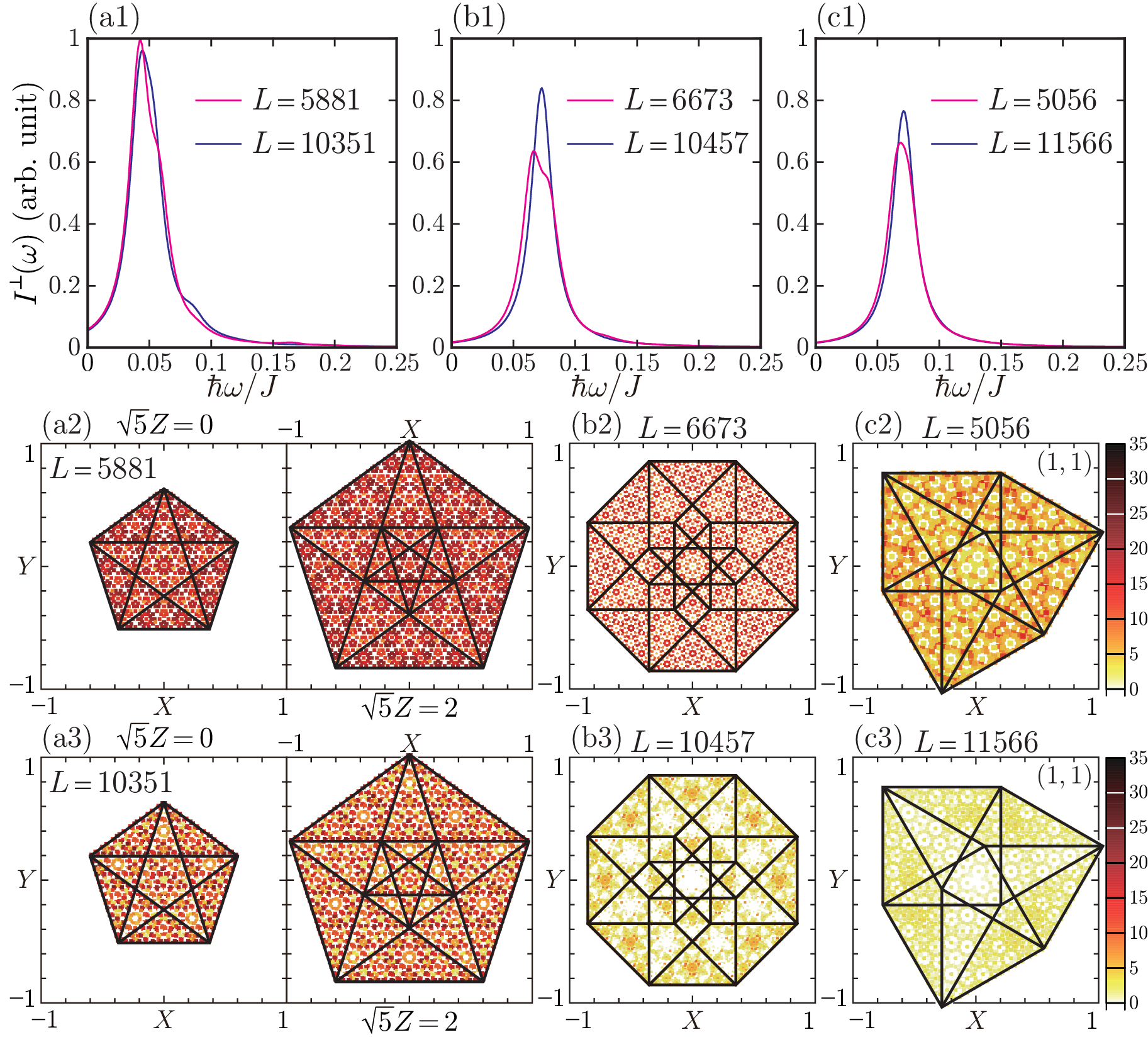}
\caption{%
         Spin-orbit mechanism magnetic Raman spectra $I^{\perp}(\omega)$
         \eqref{E:RamanIntEq} at $\varDelta=0.999$
         for the Penrose lattices of $L=5881$ and $L=10351$ (a1),
         for the Ammann-Beenker lattices of $L=6673$ and $L=10457$ (b1),
         and for the Socolar lattices of $L=5056$ and $L=11566$ (c1).
         Contour plots of site-resolved Raman spectra $I^{\perp}(\omega)|_{l}$
         \eqref{E:siteresolvedRaman} multiplied by the number of sites 
         at the peak energy: $L=5881$ (a2) and $L=10351$ (a3) for the Penrose lattices,
         $L=6673$ (b2) and $L=10457$ (b3) for the Ammann-Beenker lattices,
         and $L=5056$ (c2) and $L=11566$ (c3) for the Socolar lattices.
         }
\label{F:perpPenABSoc0.999}
\end{figure}

   Finally, we discuss the vicinity of the Heisenberg isotropic point.
Figure~\ref{F:perpPenABSoc0.999} shows the SO-mechanism Raman spectra at $\varDelta=0.999$
for the Penrose, Ammann-Beenker, and Socolar lattices, together with contour plots of
their site-resolved Raman spectra in the perpendicular space.
Reduction of the anisotropy gap leads to a single peak at low energy across all lattices.
The intensity distributions of the site-resolved Raman spectra at the peak energy exhibit
site-to-site variations without systematic dependence on the vertex type.
In terms of projection-window subclassification, this corresponds to the limiting case
in which subdomains reduce to individual sites distinguished by their distant surroundings.
With increasing system size, the peak positions remain fixed and intensities are of the
same order, while the perpendicular-space mappings show that contributions per site decrease.
This indicates the softening and itinerant character of low-energy magnons as
the system approaches the Heisenberg isotropic point \cite{W104427,Y702}.

\section{Summary and Discussion}
   We have investigated spin-orbit-interaction-driven magnetic Raman spectra of two-dimensional
bipartite-lattice \textit{XXZ} antiferromagnets with particular emphasis on a striking contrast
between quasiperiodic and periodic systems.
The SO-mechanism first-order spectra consist only of $\mathrm{A}_1$
(rotation-invariant and mirror-symmetric) species.
The exchange-mechanism second-order spectra, on the other hand, consist
only of $\mathrm{E}_2$ species \cite{I053701}.
Indeed we have $\mathrm{A}_1$ species in the exchange-mechanism fourth-order
spectra as well \cite{I053701},
but its intensity is much smaller than that in the first-order spectra.
The fourth-order intensity relative to the second-order one reads
$t^{2}/(U-\hbar\omega_{\mathrm{in}})^{2}$ with $t$, $U$, and $\omega_{\mathrm{in}}$ being
the electron hopping amplitude, on-site Coulomb repulsion, and incident light frequency,
respectively \cite{S1068,S365}.
In the far-resonant regime $t \ll |U-\hbar\omega_{\mathrm{in}}|$, the fourth-order spectra
remain sufficiently small compared to the first-order one,
even though the intensity of the first-order line amounts to one tenth of that of
the second-order peak \cite{F514}.
The ratio of the integrated intensities of single-magnon ($h_{1}$) and two-magnon ($h_{2}$)
scattering has been established for typical rutile-type antiferromagnets.
In $\mathrm{FeF_{2}}$, the ratio $h_{1}/h_{2}$ is observed to be approximately $1/3$ to $1/2$,
while in $\mathrm{MnF_{2}}$, it is estimated to be about $1/10$ \cite{F514}.
Although the SO-mechanism first-order intensity is parametrically smaller than
the exchange-mechanism second-order response, it remains well within the detectable range,
as demonstrated by these benchmarks.
There is another contrast between the first-order and second-order spectra.
The second-order spectra generally lose their intensity in the Ising limit,
reducing to elastic scatterings, whereas
the first-order spectra have their strongest intensity in the Ising limit in general,
vanishing in the Heisenberg limit.
The first-order and second-order spectra are thus complementary to track magnetic excitations
varying with exchange anisotropy.

   In the Ising limit, every first-order spectrum consists merely of delta-function peaks emergent
at $\hbar\omega=zJS$, whether the lattice is periodic or quasiperiodic.
Upon switching \textit{XY} terms on, the initial delta-function peaks split into two or more
in quasiperiodic systems, while they remain singly peaked in periodic systems.
Indeed such splittings must be the consequences of various coordination numbers $z$ coexistent,
but with which peak they begin and in what manner they grow are a mirror of each quasiperiodic
tiling manner.
Such features look more contrastive to each other in the perpendicular space.
Suppose we add $O(S^{0})$ quantum corrections to $\mathcal{H}_{\mathrm{LSW}}$
via Wick decomposition \cite{N034714,Y094412}.
The off-diagonal blocks in the $O(S^{0})$ SW  Hamiltonian read
\begin{align}
   \left[\mathbf{C}_{\mathrm{AB}}^{(0)}\right]_{i,j}
  =-l_{i,j}
   \left[
     \langle a_{i}b_{j} \rangle
    +\frac{\varDelta}{2}
     \left( \langle a_{i}^{\dagger}a_{i} \rangle
           +\langle b_{j}^{\dagger}b_{j} \rangle \right)
   \right].
\label{E:OS0correctionOffDiag}
\end{align}
The $O(S^{0})$ terms in \eqref{E:OS0correctionOffDiag} become positive in sum
in collinear antiferromagnets on bipartite lattices \cite{Y702} and effectively increase
the exchange anisotropy $\varDelta$.
Therefore, at least in the vicinity of the Ising limit, interacting SWs generally exhibit
wider splittings than LSWs.

\begin{acknowledgment}
\acknowledgment
The authors are grateful to J. Ohara for his useful comments on our numerical coding.
This work is supported by JSPS KAKENHI Grant Number 22K03502.
\end{acknowledgment}


\end{document}